\pdfoutput=1

\documentclass[aps,twocolumn,showpacs,amsmath,amssymb,nofootinbib,superscriptaddress,showkeys,prd,floatfix]{revtex4-1} 

\usepackage{epsfig}

\newcommand{\br}[1]{\langle #1\rangle}

\usepackage{xcolor}

\begin{document} 

\title{Longitudinal correlations from fluctuating strings in Pb-Pb, p-Pb, and p-p collisions}

\author{Martin Rohrmoser}
\email{mrohrmoser@ifj.edu.pl} 
\affiliation{H. Niewodnicza\'nski Institute of Nuclear Physics PAN, 31-342 Cracow, Poland}
\affiliation{Institute of Physics, Jan Kochanowski University, 25-406 Kielce, Poland}

\author{Wojciech Broniowski}
\email{Wojciech.Broniowski@ifj.edu.pl}
\affiliation{H. Niewodnicza\'nski Institute of Nuclear Physics PAN, 31-342 Cracow, Poland}
\affiliation{Institute of Physics, Jan Kochanowski University, 25-406 Kielce, Poland}

\date{26 November 2019}  

\begin{abstract}
In a framework of a semi-analytic model with longitudinally extended strings of fluctuating end-points, 
we demonstrate that the rapidity spectra and two-particle correlations in collisions
of Pb-Pb, p-Pb, and p-p at the energies of the Large Hadron Collider can be universally reproduced. 
In our approach, the strings are pulled by wounded constituents appearing in the Glauber modeling at the partonic level. 
The obtained rapidity profile for the emission of hadrons from a string yields bounds for the distributions of the 
end-point fluctuations. Then, limits for the two-particle-correlations in pseudorapidity can be obtained. 
Our results are favorably compared to recent experimental data from the ATLAS Collaboration.
\end{abstract}


\maketitle

\section{Introduction \label{sec:intro}}

Longitudinal correlations are an important source of information on the dynamics of hadronic collisions. There are numerous 
on-going efforts to understand them, both at the theoretical side, as well as in the experiments at  BNL Relativistic Heavy-Ion Collider (RHIC) and the 
CERN Large Hadron Collider (LHC), in particular with detector upgrades covering broader ranges in pseudorapidity. As is well known, the 
long-range rapidity correlations supply information on the earliest phases of the reaction, since from causality the correlations stem from 
proper times $\tau$ limited by $\tau \lesssim \tau_f \exp(-\Delta \eta/2)$, where $\tau_f$ is the freeze-out proper time and $\Delta \eta$ is the 
pseudorapidity separation of the particles in the pair. 

In our recent paper~\cite{Rohrmoser:2018shp} we have presented an analysis of the longitudinal hadronic correlations at 
the highest RHIC energy of $\sqrt{s_{\rm NN}}=200$~GeV in the framework of
a simple model, where emission proceeds from strings with fluctuating end-points~\cite{Broniowski:2015oif}. The model is, up to 
emission profiles extracted from the data, analytic, which allows us for a simple understanding of generic production features present 
in various string or fluxtube approaches. The present study provides an extension of our method to the LHC energies. 

We recall that QCD-motivated string or color flux-tube models are commonly used in particle physics phenomenology to describe the longitudinal 
dynamics. The strings extend between receding color sources and fragment, producing hadrons. 
Many sophisticated Monte Carlo codes are based on the Lund string model (see, 
e.g.,~\cite{Andersson:1983ia,Wang:1991hta,Lin:2004en,Sjostrand:2014zea,Bierlich:2018xfw,Ferreres-Sole:2018vgo}), or on the 
Dual Parton Model build on the Pomeron and Regge exchanges~\cite{Capella:1992yb,Werner:2010aa,Pierog:2013ria}. 
A common feature of these phenomenologically successful codes is a formation of 
a collection of strings pulled between the constituents of the projectiles in the early stage of the collision. 
The end-points of a string 
have opposite color charges (triplet-antitriplet for the quark-diquark and quark-antiquark configurations, or octet-octet for the gluon-gluon case). Moreover, the 
location of the string end-points in spatial rapidity $\eta_{PS}\equiv \tfrac{1}{2}\ln[(t+z)/(t-z)]$ fluctuates following a proper parton distribution function.
As argued in~\cite{Broniowski:2015oif,Rohrmoser:2018shp}, these fluctuations are the key feature enabling control over 
the one body densities (pseudorapidity spectra) and the two-particle 
correlations in pseudorapidity. In our study we focus on this effect, neglecting other features typically incorporated in Monte Carlo codes, such as 
the nuclear shadowing or baryon stopping. In our study, 
rather than using the parton distribution functions to describe the end-point distributions, we take a more flexible and phenomenological approach, where 
these distributions are adjusted to reproduce the pseudorapidity spectra. 

Another important issue is the distribution of the number of strings, which finally translates into the multiplicity of the produced hadrons. 
We use the fact that the multiplicity of the produced hadrons is successfully described within the wounded picture~\cite{Bialas:1976ed}, 
which is an adoption of the Glauber theory~\cite{Glauber:1959aa} to inelastic collisions~\cite{Czyz:1969jg}. 
Moreover, the wounded quark scaling~\cite{Bialas:1977en,Bialas:1977xp,Bialas:1978ze,Anisovich:1977av} has been shown 
to work surprisingly well~\cite{Eremin:2003qn,KumarNetrakanti:2004ym,Bialas:2006kw,Bialas:2007eg,Alver:2008aq,%
Agakishiev:2011eq,Adler:2013aqf,Loizides:2014vua,Adare:2015bua,%
Lacey:2016hqy,Bozek:2016kpf,Zheng:2016nxx,Sarkisyan:2016dzo,Mitchell:2016jio,Chaturvedi:2016ctn,Loizides:2016djv,Tannenbaum:2017ixt,Barej:2017kcw,Barej:2019xef} at both
RHIC and the LHC collision energies. Extensions to more partons per nucleon than just three quarks have also been considered, with the conclusion that 
the increase in energy yields more wounded partons~\cite{Bozek:2016kpf}. In the present study we use the wounded model with a few (3 to 6) constituents per nucleon.

\section{The model}

As mentioned, our model combines the string picture with the wounded parton model, assuming that the number of strings is given by the number of the wounded constituents. 
As a matter of fact, this complies to the Lund model mechanism, where the basic string extends between a parton from a given nucleon and a parton (or diquark) 
from {\em the same} nucleon~\cite{Andersson:1983ia}.
Thus, in collisions of nuclei $A$ and $B$, hadrons are emitted from strings associated to mutually independent $N_A$ wounded partons
from $A$ and $N_B$ wounded partons from $B$, respectively. At a given collision energy 
the emission profile of hadrons (defined as the number of hadrons per $\eta$) from each string, $f(\eta)$, is assumed to be universal, i.e., independent 
of the mass numbers of the projectiles or centrality.
Here $\eta$ denotes the pseudorapidity in the center of mass of the colliding NN system. 
The above assumptions correspond to the following scaling law~\cite{Bialas:2004su}:
\begin{equation}
\frac{dN_{\rm ch}}{d\eta}=\langle N_A \rangle f(\eta)+ \langle N_B\rangle f(-\eta),
\label{eq:wqm}
\end{equation}
where we have adopted the convention that $A$ moves to the right and $B$ to the left
along the $z$ axis.
The symbol  $\langle . \rangle$ denotes the average over events in the considered centrality class.

From Eq.~(\ref{eq:wqm}) it follows that the symmetric and antisymmetric parts of the distributions are given by
\begin{eqnarray}
\frac{1}{2}\left(\frac{dN}{d\eta}(\eta)+\frac{dN}{d\eta}(-\eta)\right)=\langle N_+\rangle f_s(\eta), \nonumber \\
\frac{1}{2}\left(\frac{dN}{d\eta}(\eta)-\frac{dN}{d\eta}(-\eta)\right)=\langle N_-\rangle f_a(\eta),
\label{eq:sym}
\end{eqnarray}
with $N_\pm=N_A \pm N_B$ denoting the sum and the difference of sources from $A$ and $B$, whereas $f_s(\eta)$ and $f_a(\eta)$
denote the symmetric and antisymmetric parts of the profile $f(\eta)$.

In our simulations, centrality is determined via the quantiles of the total number of wounded partons, $\langle N_+ \rangle$.

From various studies of hadron multiplicity distributions in p-Pb collisions, it is known that the Glauber approach 
of hadron production must be amended with 
fluctuations of the number of sources. Typically, the negative binomial distribution is overlaid over the distribution of wounded sources. 
We follow this scheme in our simulations, with the following prescription: we generate events with {\tt GLISSANDO~3}~\cite{Bozek:2019wyr}, with $n_A$ and $n_B$ 
wounded partons in a given event. Then we generate randomly $N_A=k(N_A;n_A,q)$ and $N_B=k(N_B;n_B,q)$, where 
\begin{eqnarray}
k(x;n,q)={\rm NB} \left [x;\frac{n q}{1 - q}, q \right ] \label{eq:nb}
\end{eqnarray} 
is the negative binomial distribution with $x=0$ removed, i.e., $x=1,2,3,\dots$. 
The cases where $N_A=0$ or $N_B=0$ (no strings) are disregarded.
By construction, $\langle x \rangle=n$ and  ${\rm var}(x)= n/q$. 
The parameter $q \le 1$, treated as a free variable to be fitted, controls the variance of the number of strings. 

The role of increased fluctuations introduced by an overlaid distribution enters indirectly into our 
analysis, by modifying the division of the event sample into centrality classes.  

\section{Extraction of the emission profile from pseudorapidity spectra \label{sec:prof}}

\begin{figure*}[tb]
\centering
\includegraphics[clip=true,trim=10 7 0 0,width=0.4\textwidth]{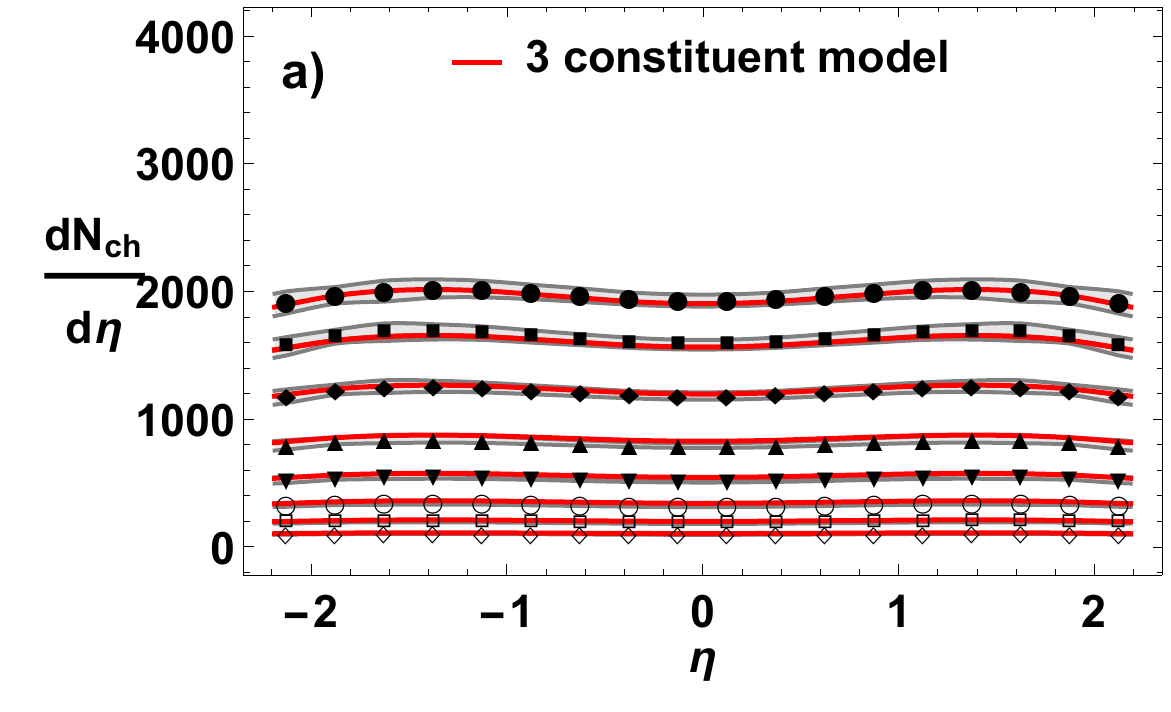}
\includegraphics[clip=true,trim=10 7 0 0,width=0.4\textwidth]{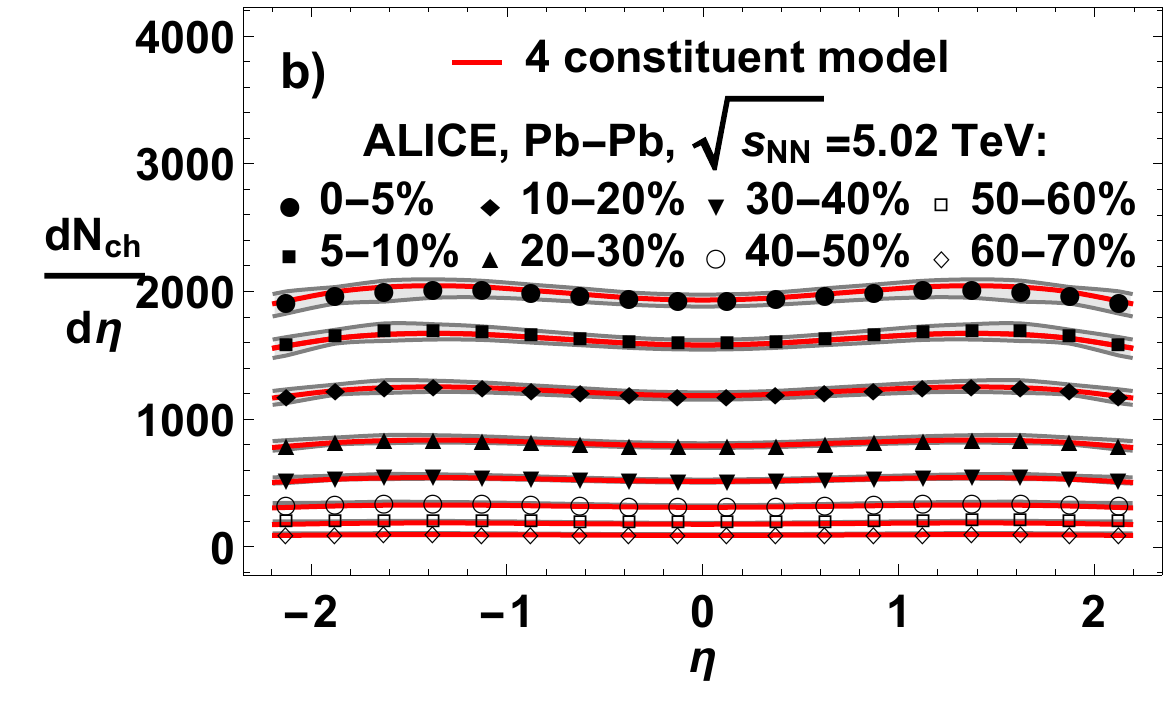}\\
\includegraphics[clip=true,trim=10 7 0 0,width=0.4\textwidth]{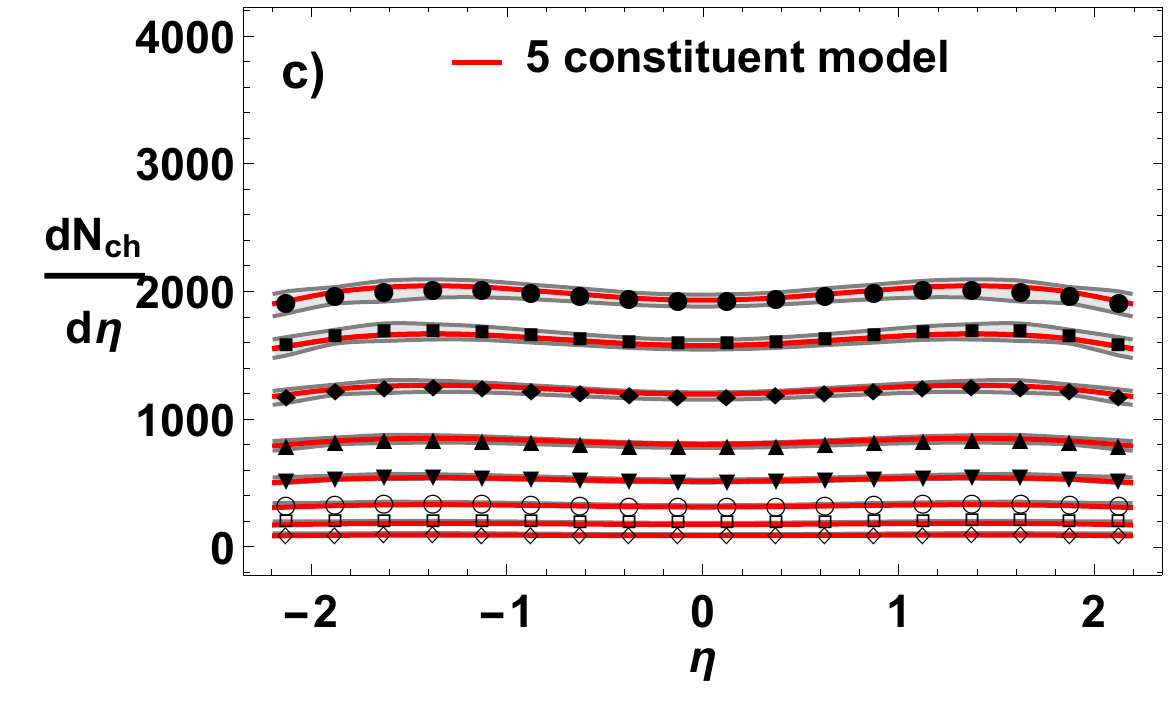}
\includegraphics[clip=true,trim=10 7 0 0,width=0.4\textwidth]{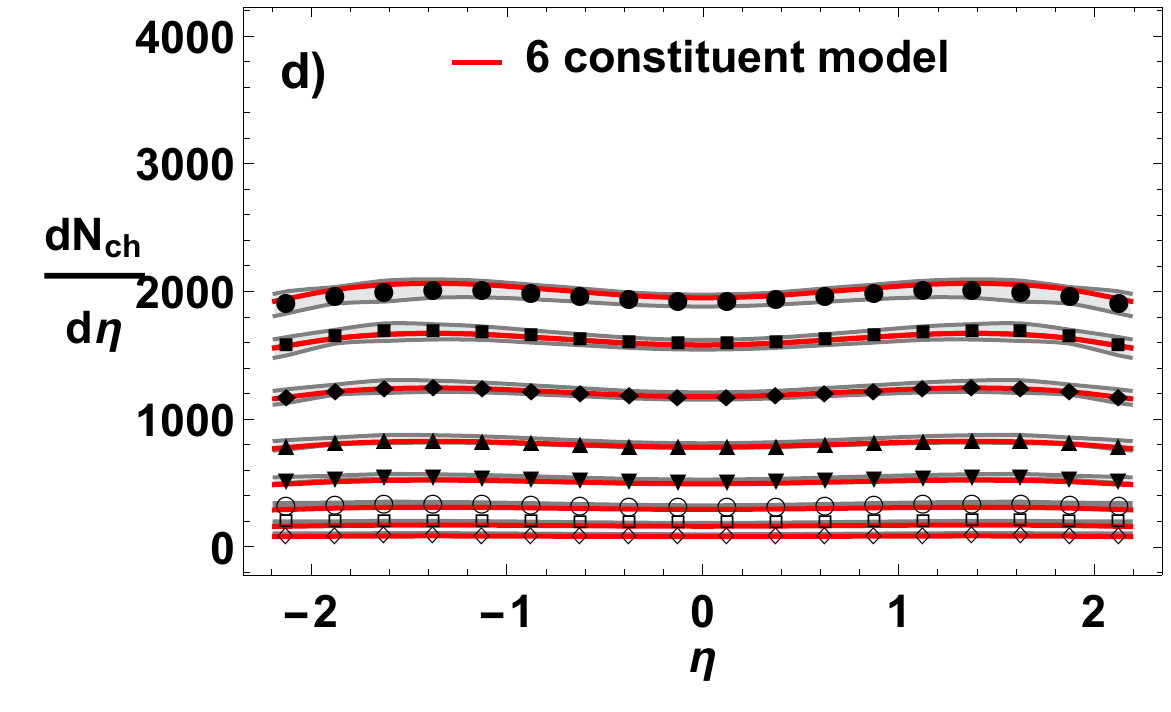}
\caption{Model results (solid lines) and ALICE data~\cite{Adam:2016ddh} (points with bands indicating experimental errors) 
for the pseudorapidity spectra in Pb-Pb collisions at $5.02$~TeV. Subsequent panels correspond to models with 3, 4, 5, and 6 partons per nucleon.
\label{fig:PbPb}}
\end{figure*}

\begin{figure*}[tb]
\centering
\includegraphics[clip=true,trim=9 7 0 0,width=0.4\textwidth]{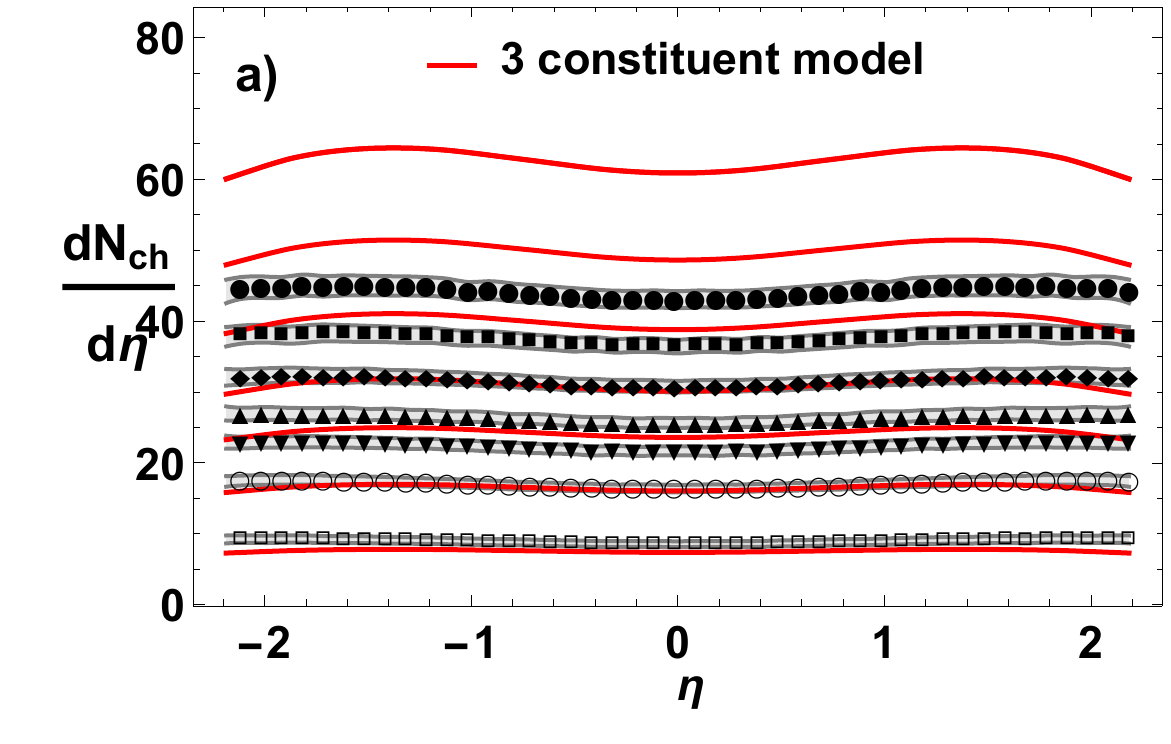}
\includegraphics[clip=true,trim=9 7 0 0,width=0.4\textwidth]{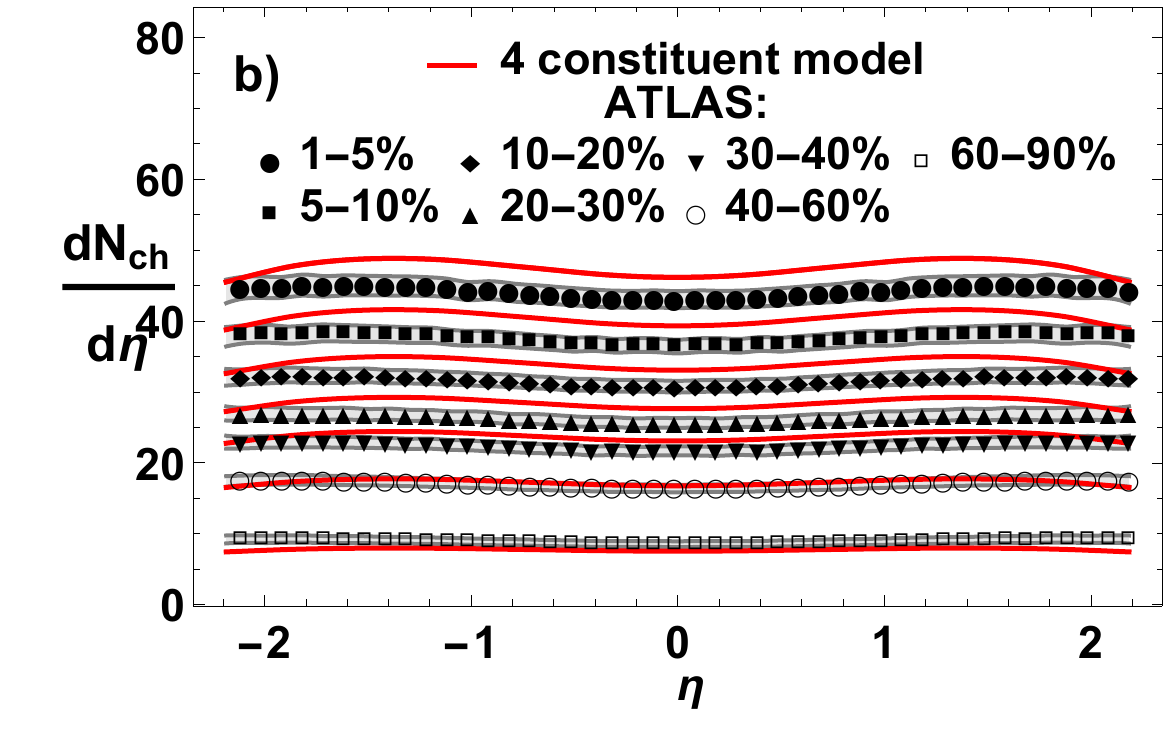}\\
\includegraphics[clip=true,trim=9 7 0 0,width=0.4\textwidth]{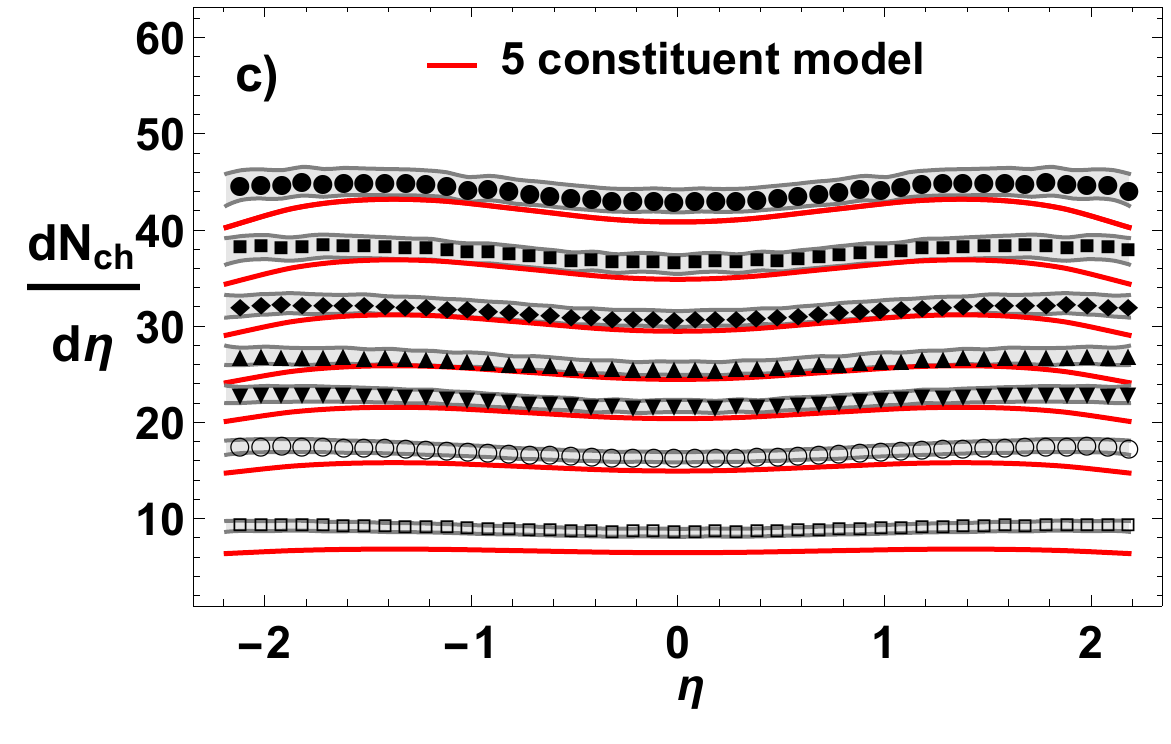}
\includegraphics[clip=true,trim=9 7 0 0,width=0.4\textwidth]{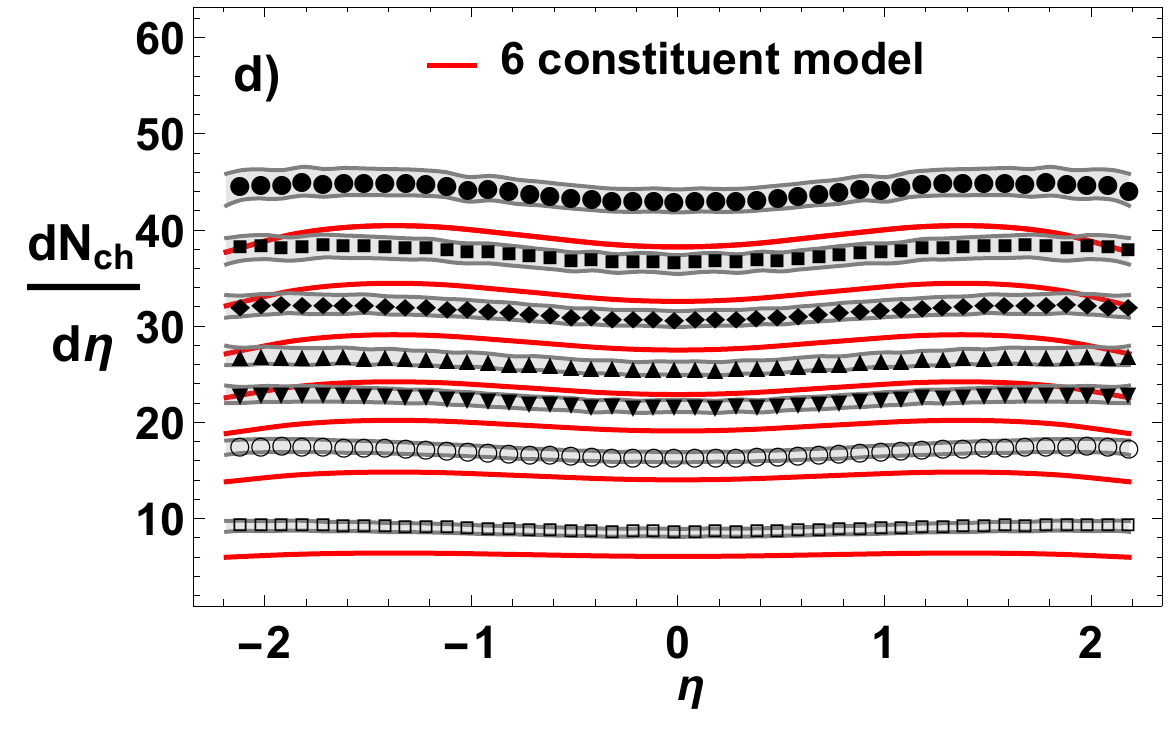}
\caption{Model results (solid lines) and ATLAS data~\cite{Aad:2015zza} (points with bands indicating experimental errors) 
for the symmetric parts of the pseudorapidity spectra in p-Pb collisions at $5.02$~TeV. Subsequent panels correspond to models with 3, 4, 5, and 6 partons per nucleon.}
\label{fig:pPbsym}
\end{figure*}

\begin{figure*}[tb]
\centering
\includegraphics[clip=true,trim=10 7 0 0,width=0.4\textwidth]{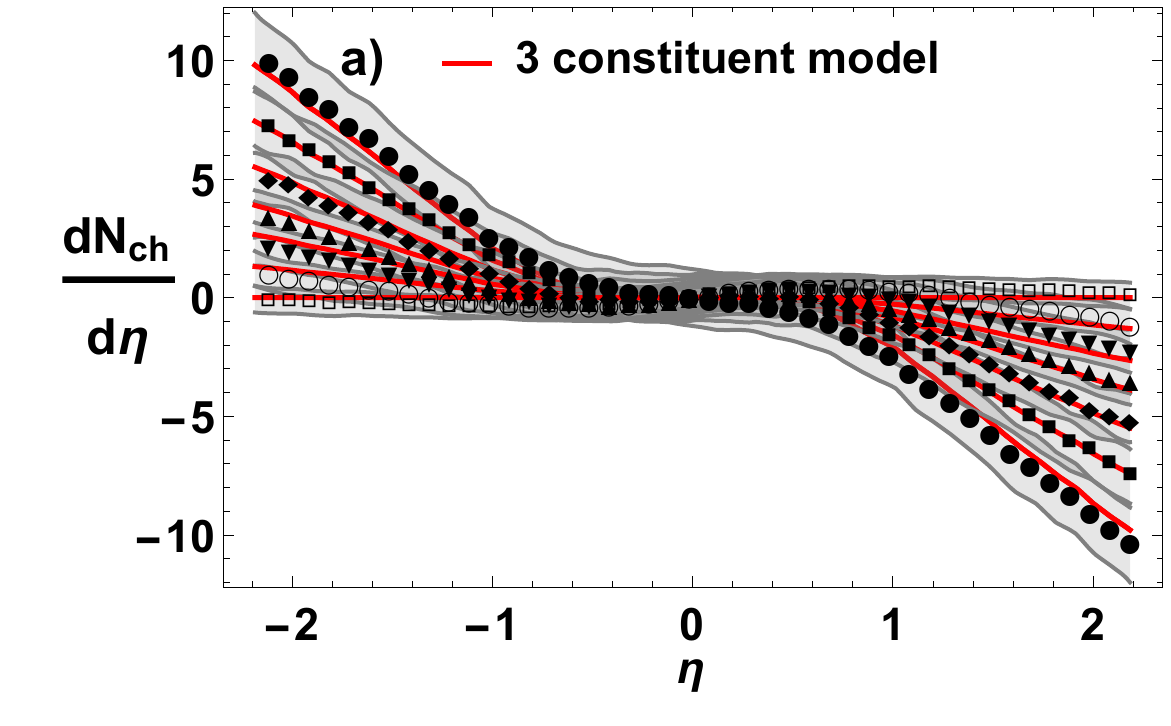}
\includegraphics[clip=true,trim=10 7 0 0,width=0.4\textwidth]{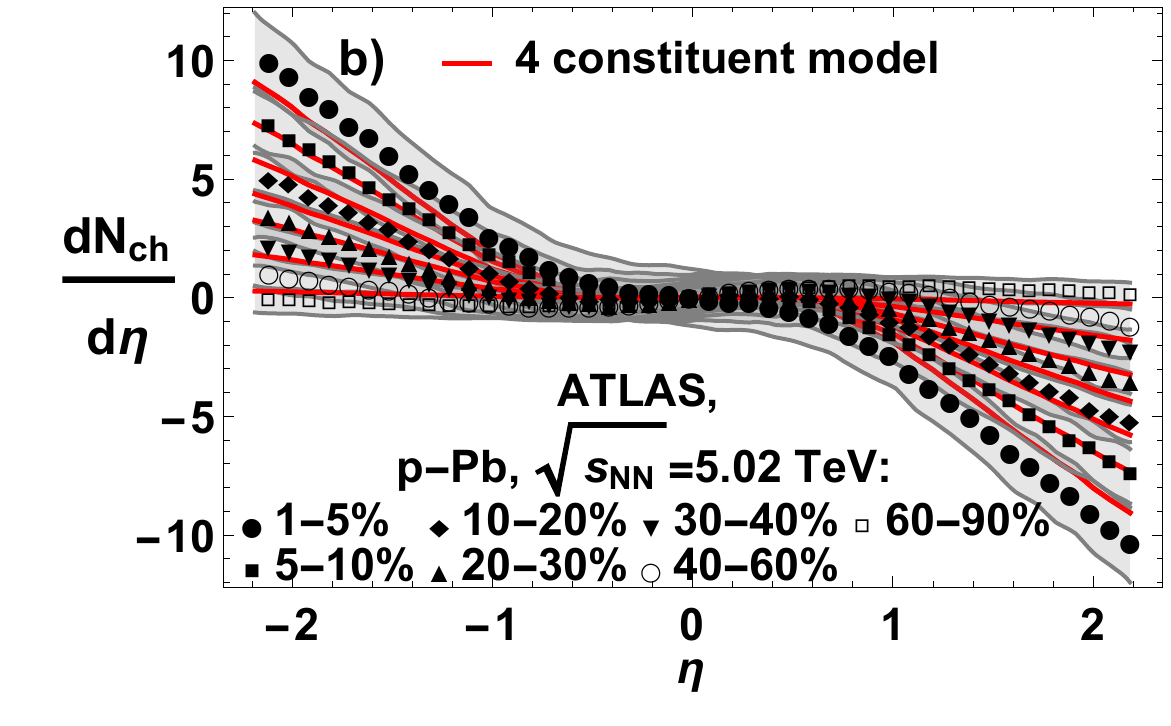}\\
\includegraphics[clip=true,trim=10 7 0 0,width=0.4\textwidth]{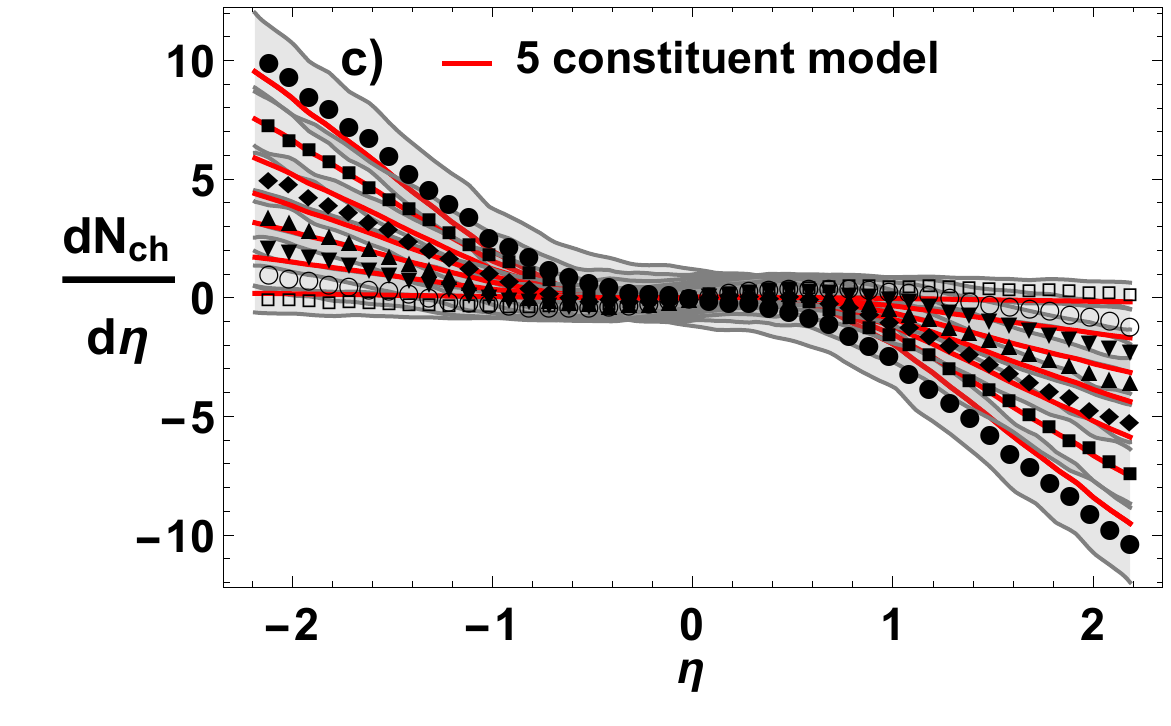}
\includegraphics[clip=true,trim=10 7 0 0,width=0.4\textwidth]{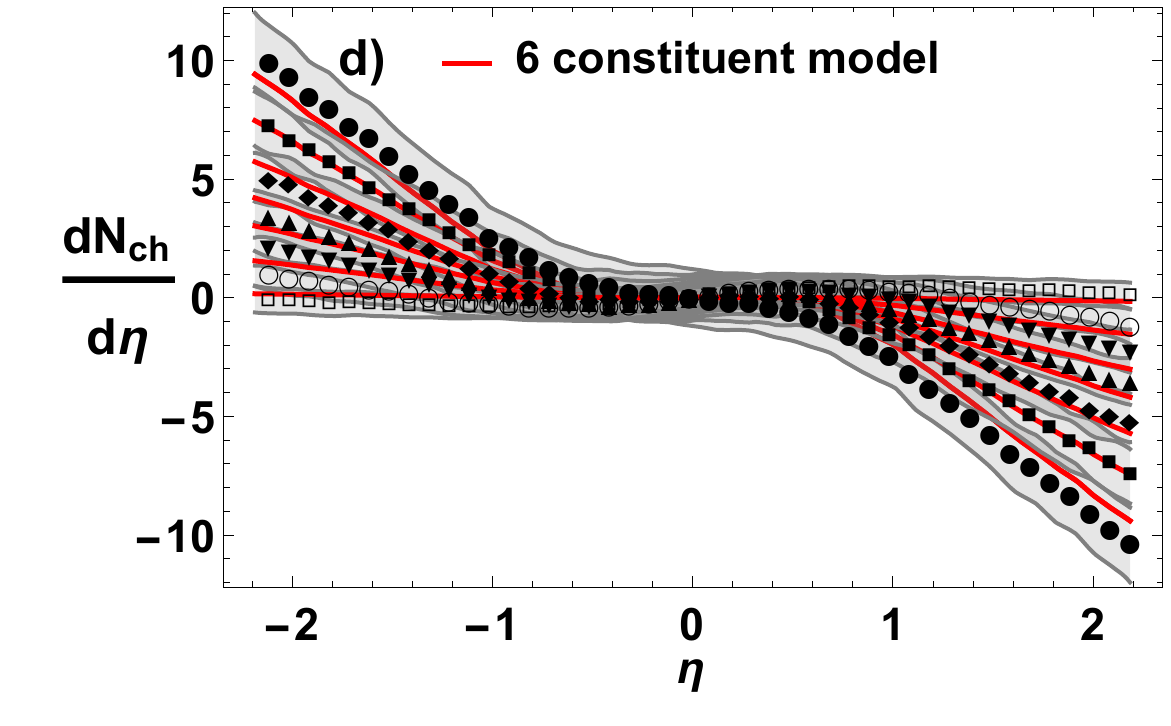}
\caption{Same as in Fig.~\ref{fig:pPbsym}, but for the antisymmetric part of the p-Pb spectra.}
\label{fig:pPbasym}
\end{figure*}

This section explains in detail how the experimental data on rapidity spectra from  p-Pb and Pb-Pb
collisions at the LHC with $\sqrt{s_{\rm NN}}=5.02$~TeV were jointly fitted to obtain the emission profiles of Eq.~(\ref{eq:wqm}).
We carry out the following steps:
\begin{enumerate}
\item Choose the variant of the model by deciding on the number of partons per nucleon.
\item {\tt GLISSANDO~3} is run to generate event samples with the number of wounded sources $n_A$ and $n_B$  for both the p-Pb and Pb-Pb collisions. 
\item For a given value of $q$ samples with string numbers $N_A$ and $N_B$ are obtained by overlaying a negative binomial 
distribution according to Eq.~(\ref{eq:nb}).
\item The samples for p-Pb and Pb-Pb are then divided into centrality classes according to the values of $N_+$.
\item The values of $\langle N_A \rangle$ and $\langle N_B \rangle$ in all centrality classes are used to construct the least squares fit of the profiles $f(\eta)$ 
to the experimental pseudorapidity spectra ${dN_{\rm ch}}/{d\eta}$ from Eq.~(\ref{eq:wqm}).
\item Steps 3-5 are repeated to obtain the optimum value of $q$, 
corresponding to a global minimum of the least square function. The result is the optimum profile $f(\eta)$.
\end{enumerate}

In the fitting procedure, the symmetric part of the profile, $f_s(\eta)$, is sensitive to both the Pb-Pb and p-Pb data, 
whereas the the antisymmetric part,  $f_a(\eta)$, depends only on the (asymmetric) p-Pb collisions, as is obvious from Eq.~(\ref{eq:sym})  
As we wish to fit jointly the Pb-Pb and p-Pb, we choose the collision energy where both sets of data on the pseudorapidity 
spectra are available, namely $\sqrt{s_{NN}}=5.02$~TeV. Specifically, we take ALICE data~\cite{Adam:2016ddh} on pseudorapidity spectra of Pb-Pb, and the 
ATLAS data~\cite{Aad:2015zza} on pseudorapidity spectra of p-Pb. We 
note that the used data for the p-Pb pseudorapidity spectra for p-Pb collisions from ATLAS~\cite{Aad:2015zza} are accurately compatible
to the ALICE~\cite{Adam:2016ddh} data with the V0A selection of centrality. We prefer to use the ATLAS data here, as we will compare 
the results of our model with the pseudo-rapidity correlations extracted from the ATLAS-experiment~\cite{Aad:2015zza}.

The experimental data for p-Pb collisions at the LHC are shifted with respect to the NN center-of-mass frame by 0.465 units of rapidity. 
Since rapidity $y$ and pseudorapidity $\eta$ are related as $p_T \sinh(\eta)=\sqrt{m^2+p_T^2}\sinh(y)$, where $p_T$ is the 
transverse momentum and $m$ the particle's mass, one can obtain $\eta\approx y$ in the case $m\ll p_T$, which we assume in our further considerations. 
One can justify this assumption by the fact that the emitted particles are predominantly pions with a small rest mass of $m_\pi \simeq 140$~MeV, which is smaller 
than typical values of $p_T$.
Thus, to a good approximation the pseudorapidity $\eta$ in the lab frame is related to $\eta$ in the NN center-of-mass frame, 
$\eta_{\rm lab}\simeq \eta_{\rm CM}+0.465$. This allows us to transform the experimental pseudorapidity spectra 
from p-Pb collisions into the CM frame by a simple shift.

A joint least squares fit for Pb-Pb and p-Pb spectra can be performed in the following way:
For each value of $q$ individually the numbers of sources in the wounded parton model, overlaid with the negative binomial distribution, are generated with the 
help of {\tt GLISSANDO~3}~\cite{Bozek:2019wyr}. 
In order to obtain the emission profile $f(\eta)$ 
we construct for each value of $\eta$ for which the data exists the least squares sum $L(f_A(\eta),f_B(\eta))$, depending on two fitting parameters $f_A(\eta)$ and $f_B(\eta)$,
\begin{eqnarray}
&&  L(f_A(\eta),f_B(\eta))= \label{eq:lss} \\ 
&&  \frac{1}{N}\sum_{i=1}^N \left\{ \left[ \left(\frac{dN}{d\eta}(\eta)\right)_i-(\langle N_A \rangle_i f_A(\eta)+ \langle N_B\rangle_i f_B(\eta))\right]^2 \right . \nonumber \\
&& +\left . \left[ \left(\frac{dN}{d\eta}(-\eta)\right)_i-(\langle N_B \rangle_i f_A(\eta)+ \langle N_A\rangle_i f_B(\eta))\right]^2 \right\}, \nonumber
\end{eqnarray}
where $i$ runs over all spectra (i.e., all the centrality classes and reactions) 
that are to be fitted\footnote{We use all the available rapidity spectra for Pb-Pb and p-Pb, except the p-Pb data for the most central $1\%$ of collisions, which are far of he optimal fit, 
hinting different physics in this case.}. 
We then minimize $L(\eta,f_A,f_B)$ at each $\eta$, which yields the functions $f_A(\eta)$ and $f_B(\eta)$.
Our choice for the least squares sum, Eq.~(\ref{eq:lss}), has the desired symmetry property $L(-\eta,f_A,f_B)=L(\eta,f_B,f_A)$, 
which follows from the fact that $f_A(\eta)=f_B(-\eta)$, which means the replacement of the left-going wounded source by the right-going one.  
Recall that in the notation of Eq.~(\ref{eq:wqm})   $f(\eta)=f_A(\eta)=f_B(-\eta)$.

\begin{table}[b]
\caption{Optimum values of the negative binomial parameter $q$ and the corresponding value of the least squares function $\hat{L}$, for models with 
various numbers of partons per nucleon. \label{tab:q}}
\begin{center}
\begin{tabular}{c|c|c}
constituents&$q$&$\tilde{L}$\\\hline
3&0.245&476\\
4&0.905&140\\
5&0.785&137\\
6&0.805&571\\
\end{tabular}
\end{center}
\end{table}

The procedure described above provides the optimum emission spectrum $f(\eta)$ for a given value of the negative binomial parameter $q$ of Eq.~(\ref{eq:nb}) . 
To obtain the optimum value of $q$ we additionally minimize  the least squares sum (\ref{eq:lss}) summed over all values of $\eta$, denoted as $\hat{L}$,
with respect to $q$.
The optimum values for $q$ for the models with 3, 4, 5, and 6 partons per nucleon
together with the corresponding value for the least squares sum $\tilde{L}$ are listed in Table~\ref{tab:q}.
We note that the values for $\tilde{L}$ are lowest 
for models with $4$ or $5$ constituents per nucleon. Thus in the following we focus on results for these two cases.

The results of our fits for the symmetric parts of the Pb-Pb pseudorapidity spectra for the models with 3, 4, 5, and 6 partons per nucleon are shown in Fig.~\ref{fig:PbPb}.
As the figure shows, the ALICE data~\cite{Adam:2016ddh} are reasonably well reproduced for all variants of the model and for all centrality selections.
Thus the Pb-Pb spectra do not discriminate between the variants of the model.  

The situation is different for the p-Pb case.
Figures~\ref{fig:pPbsym} and~\ref{fig:pPbasym} show, correspondingly, the symmetric and antisymmetric parts 
of the pseudorapidity spectra for p-Pb collisions, compared the ATLAS data~\cite{Aad:2015zza}.
Whereas for the antisymmetric contributions, shown in Fig.~\ref{fig:pPbasym}, all variants of the model 
reproduce the data reasonably well, significant differences can be noticed in the symmetric contributions, shown in Fig.~\ref{fig:pPbsym}.
Acceptable agreement is obtained for $4$ and $5$ partons. In the following parts of this article, when considering correlations, we will thus 
focus on the $4$ and $5$ parton cases. 

\begin{figure}
\centering
\includegraphics[width=0.46\textwidth]{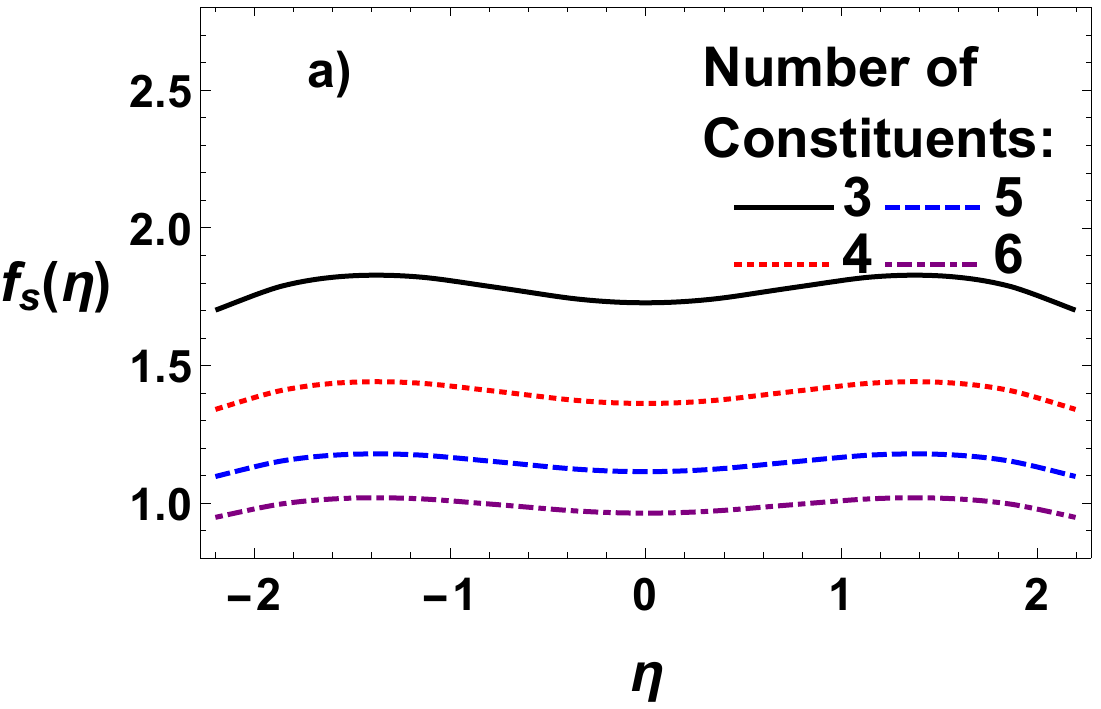}
\includegraphics[width=0.46\textwidth]{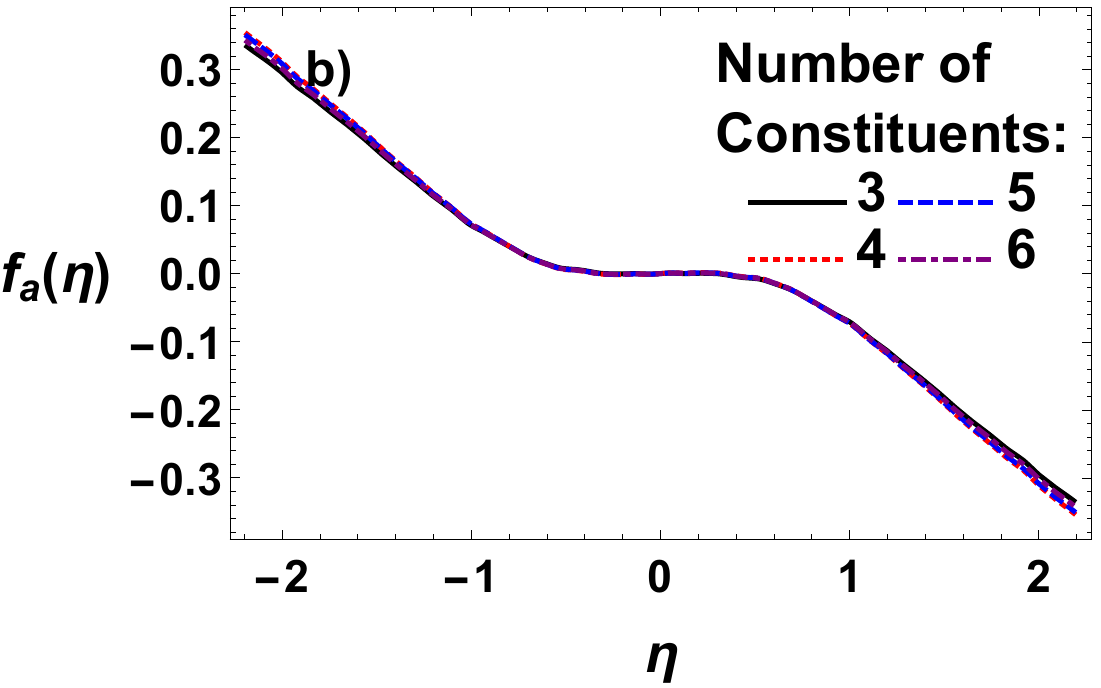}
\caption{Symmetric (a) and antisymmetric (b) parts of the emission profile $f(\eta)$  for models with various number of constituents per nucleon.} 
\label{fig:fsaetasa}
\end{figure}

The corresponding universal profiles $f(\eta)$ obtained from our fitting procedure are shown in Fig.~\ref{fig:fsaeta},
together with their symmetric and antisymmetric contributions $f_s(\eta)$ and $f_a(\eta)$ given in Fig.~\ref{fig:fsaetasa}.
We note from Fig.~\ref{fig:fsaeta} that the profiles scaled by the central value, $f(\eta)/f(0)$, differ by 
a few percent at peripheral values of $\eta$, with a steeper the fall-off with $\eta$ for larger number of partons per nucleon.

Figure~\ref{fig:fsaetasa}a) shows that the symmetric parts of the profiles, $f_s(\eta)$, decrease with the number of partons. 
This behavior is natural and follows from the first of Eq.~(\ref{eq:sym}).
When $\langle N_+\rangle$ decreases due to a smaller number of partons per nucleon, the magnitude of $f_s(\eta)$ needs to be 
correspondingly increased to yield the same pseudorapidity spectra. 
Figure~\ref{fig:fsaetasa}b) shows the antisymmetric parts of the profiles,  $f_a(\eta)$. As can be seen, 
the different number of partons per nucleon has essentially no influence on the $f_a(\eta)$.
We have found no apparent physical reason for such a behavior, which may be considered accidental.
The overall steeper fall-off of profiles $f(\eta)/f(0)$ in Fig.~\ref{fig:fsaeta} with increasing number of partons per nucleon  
can thus be understood via the decrease of the magnitude of $f_s(\eta)$, with no 
change in $f_a(\eta)$.

We remark that instead of the least squares sum of Eq.~(\ref{eq:lss}) we can use the $\chi^2$ function, which yields essentially the same optimum results. 
As to the values of $\chi^2$/d.o.f., admittedly they are large due to the approximate nature of our model, which assumes a very simple 
uniform mechanism of string production and breaking. Thus the values of $\chi^2$/d.o.f. cannot be used as stringent measures of the statistical 
quality of the fit, which is an issue shared by many models applied to ultra-relativistic nuclear collisions.

To conclude this section, as a preliminary step of our study we were able to uniformly 
fit in an approximate way 
the experimental data for Pb-Pb and p-Pb collision from the ALICE~\cite{Adam:2016ddh} and ATLAS~\cite{Aad:2015zza} collaborations, respectively, 
in the wounded parton model, with 
a preference for a model variant with 4 or 5 wounded constituents per nucleon.

\section{String end point distributions \label{sec:str}}

\begin{figure}
\centering
\includegraphics[width=0.46\textwidth]{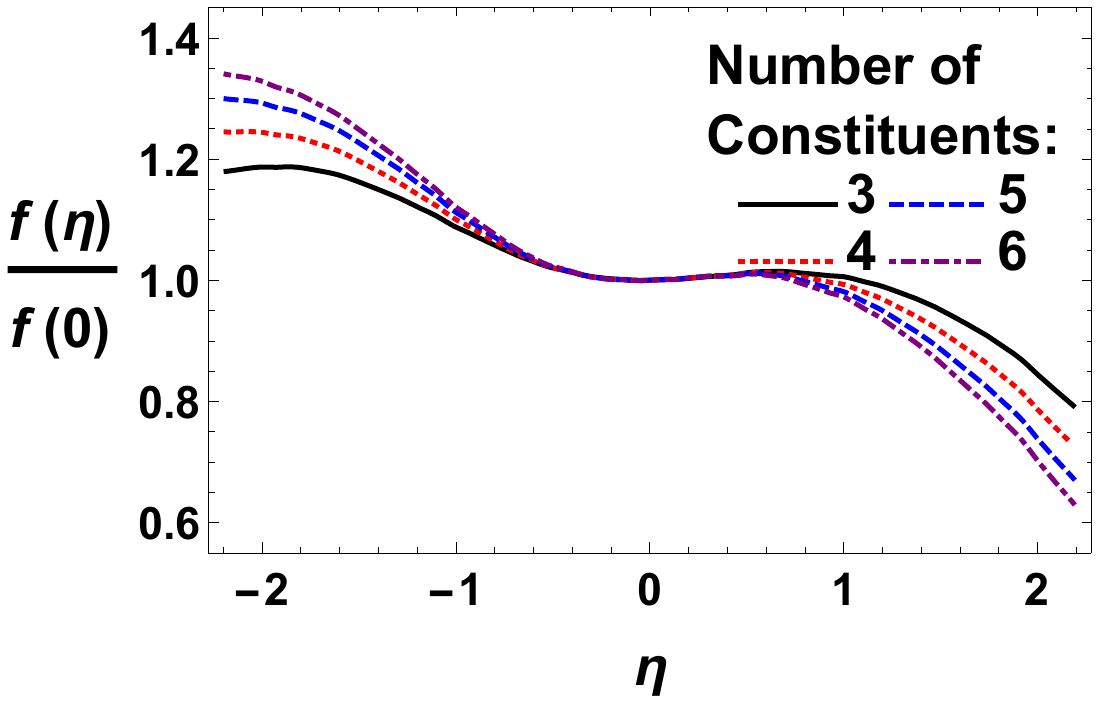}
\caption{Emission profile in pseudorapidity, divided by its value at the origin, $f(\eta)/f(0)$, for models with various number of constituents per nucleon.}
\label{fig:fsaeta}
\end{figure}

In this section we proceed in analogy to our earlier work~\cite{Rohrmoser:2018shp}.
However, in contrast to the description used therein, in this article we pass from the profile
functions in pseudorapidity $\eta$, obtained in the previous section, to the profile functions in rapidity $y$.
The reason is technical but relevant. The method of~\cite{Rohrmoser:2018shp} works for profile functions with are unimodal (have a single maximum), as this is 
what follows from strings continuously stretched between fluctuating end-points. 
Unimodality is not the case in the present analysis, as can be seen from Fig.~\ref{fig:fsaeta}. For instance, for 
the case of three constituent partons per nucleon, one can notice a maximum at $\eta\simeq -1.8$ and another weak maximum at $\eta\simeq 0.8$ 
(variants with a larger number of constituents have a maximum outside of the left bound of the plot). However, the maximum near 0.8 is an artifact of using 
pseudorapidity rather than rapidity.

As can be seen from the experimental data~\cite{Back:2001xy, Adam:2014qja,Aad:2015zza,Adam:2016ddh}, pseudorapidity 
spectra mainly differ from rapidity spectra by a pronounced dip around $\eta=0$, which trivially follows from the kinematic relation between rapidity and pseudorapidity. 
In order to pass from pseudorapidity to rapidity for the spectra which are largely dominated by the pions, 
we use the simplifying assumption of a factorization of the rapidity and $p_T$ dependence of the spectra. Then, approximately, one can write
\begin{align}
\frac{dN}{dy}=\int dp_T \frac{d\eta (y, p_T, m_\pi)}{dp_T dy}\frac{dN}{d\eta}\approx \left.\frac{d\eta}{dy}\frac{dN}{d\eta}\right|_{y\approx\eta}\,.
\label{eq:etatoy}
\end{align}
The Jacobian ${d\eta}/{dy}$ in the last part of Eq.~(\ref{eq:etatoy}) can be obtained from the experimental data from ALICE for 
the $5\%$ most central Pb-Pb collisions~\cite{Adam:2016ddh}, where both the rapidity, $dN/dy$, and pseudorapidity, $dN/d\eta$, spectra are provided.
This procedure, in essence, is a way of averaging over the transverse momentum $p_T$, incorporating the experimental acceptance.

\begin{figure}
\centering
\includegraphics[clip=true,trim=25 0 0 0,width=0.46\textwidth]{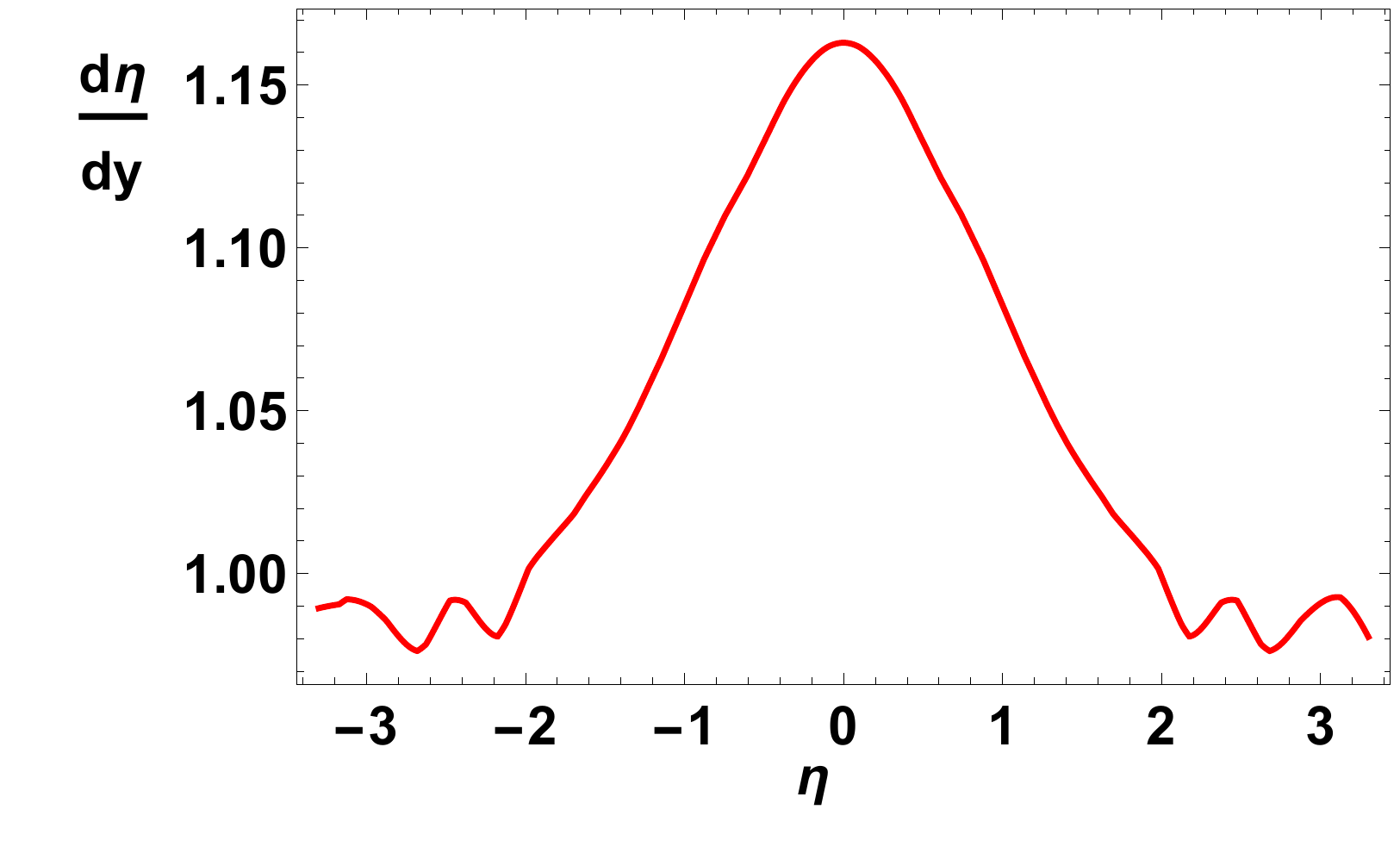}
\caption{The Jacobian $d\eta/dy$ obtained from Eq.~\ref{eq:yeta} and the experimental data from~\cite{Adam:2016ddh}.} 
\label{fig:dedy}
\end{figure}

Consequently, we can obtain the one-body emission profiles $f(y)$ in terms of $f(\eta)$ presented in Sec.~\ref{sec:prof}, namely,
\begin{eqnarray}
f(y)= \frac{d\eta}{dy} f(\eta). \label{eq:yeta}
\end{eqnarray}
Thus obtained result for $d\eta/dy$ is shown in Fig.~\ref{fig:dedy}.
Similarly, for the two-particle emission profiles we get 
\begin{eqnarray}
f_2(y_1,y_2)= \frac{d\eta_1}{dy_1} \frac{d\eta_2}{dy_2} f_2(\eta_1,\eta_2),  \label{eq:yeta2}
\end{eqnarray}
which will be used in the next section.

\begin{figure}
\includegraphics[width=0.46\textwidth]{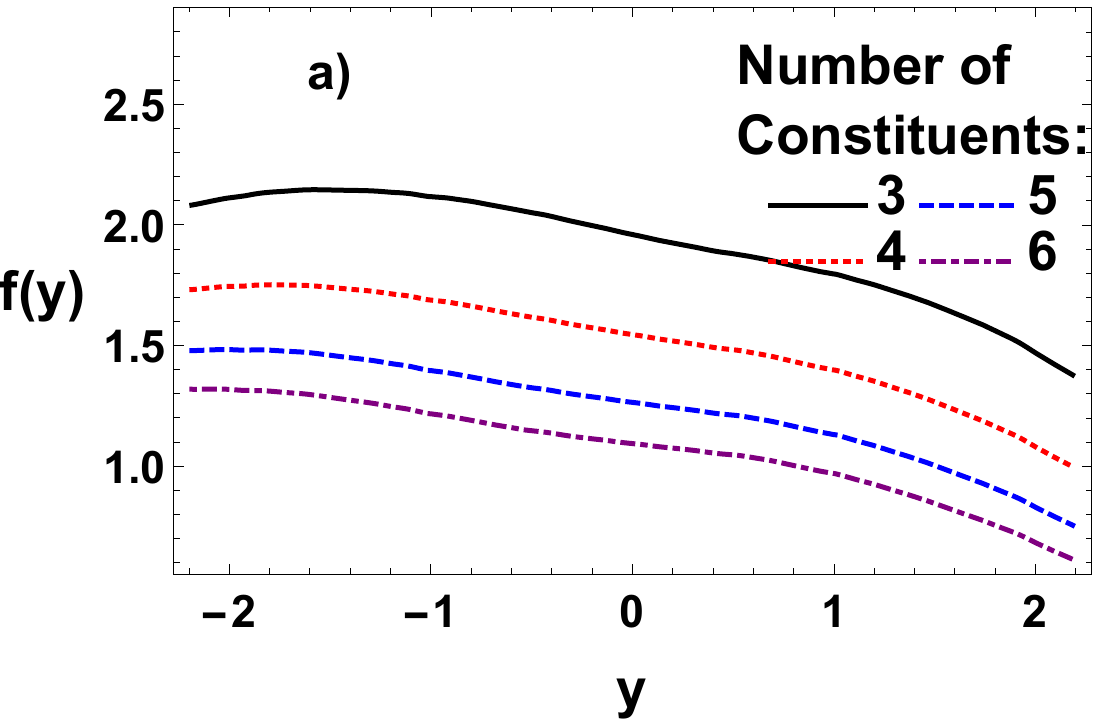}
\includegraphics[width=0.46\textwidth]{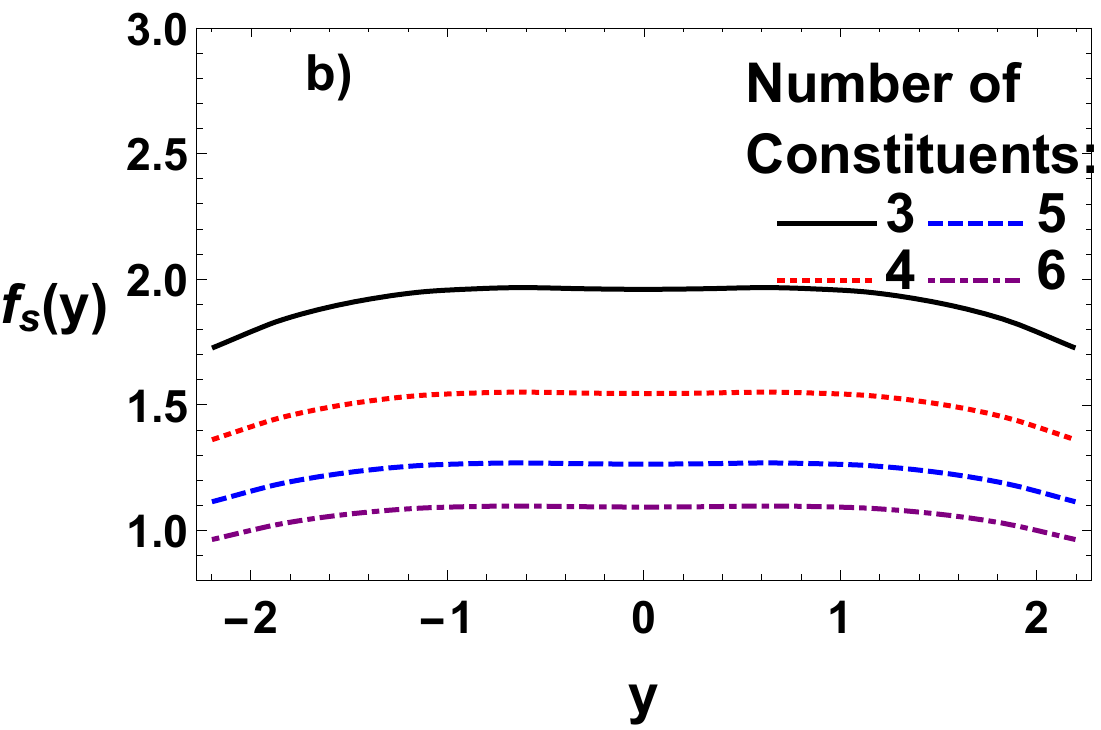}
\includegraphics[width=0.46\textwidth]{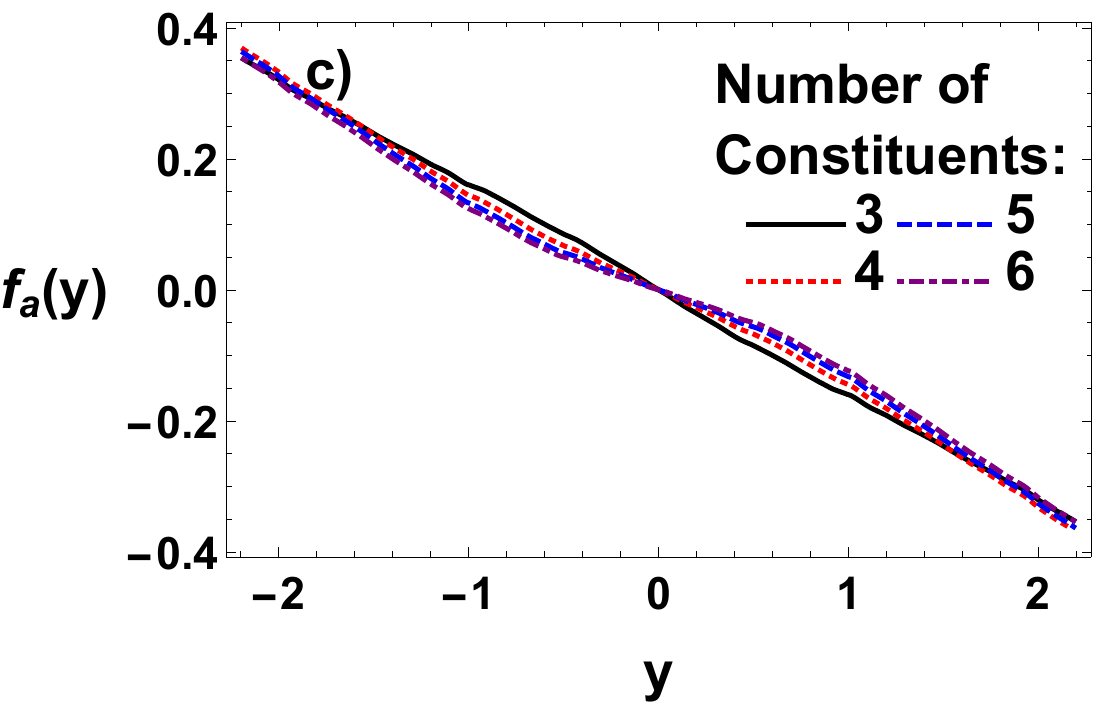}
\caption{Emission profiles of individual strings in rapidity for models with various numbers of wounded constituents (a) together with their respective symmetric (b) and antisymmetric parts (c).}
\label{fig:profiley}
\end{figure}

The results for $f(y)$ in models with 3, 4, 5, and 6 wounded partons obtained with Eq.~(\ref{eq:yeta}) are shown in Fig.~\ref{fig:profiley}. 
The feature that can be seen when comparing to $f(\eta)$ from Fig.~\ref{fig:fsaeta} is the absence of the central dip in the symmetric part. As a result, $f(y)$ 
at various centralities are unimodal functions (have only one maximum at negative $y$), which allows to carry out the analysis along the 
lines of~\cite{Rohrmoser:2018shp}. We recapitulate the basic steps of the procedure: 
\begin{enumerate}
\item Each of the $N_A$ and $N_B$ wounded sources is associated to a longitudinally extended string.
\item A string breaking at spatial rapidity $y$ corresponds to a particle emission at rapidity $y$. 
The corresponding probability distribution for string breaking, $s(y;y_1,y_2)$,  
is uniform between the end-points $y_1$ and $y_2$, namely $s(y;y_1,y_2)=\omega (\theta(y_1<y<y_2)+\theta(y_2<y<y_1))$, 
where $\omega$ is a normalization constant and the function $\theta$ equals $1$ wherever the condition in its argument is fulfilled, and $0$ otherwise.
\item String-end points $y_1$ and $y_2$ follow distributions $g_1(y_1)$ and $g_2(y_2)$, respectively. 
The corresponding cumulative distribution functions (CDFs) are denoted as $G_1(y_1)$ and $G_2(y_2)$.
\end{enumerate}

Then,  the one-body emission profile $f(y)$ can be written as~\cite{Rohrmoser:2018shp}
\begin{align}
f(y)&=\int_{-\infty}^\infty dy_1 \int_{-\infty}^\infty dy_2 g_1(y_1)g_2(y_2)s(y;y_1,y_2)\nonumber\\
&= \omega \left \{ \tfrac{1}{2} - 2 [G_1(y)-\tfrac{1}{2}]   [G_2(y)-\tfrac{1}{2}]    \right \}.
\label{eq:fg1g2}
\end{align}
It is apparent from Eq.~(\ref{eq:fg1g2}) that for a given one-body emission profile the solution to the string-end-point distributions $G_1(y)$ and $G_2(y)$ 
are not unique.
It is, nevertheless, possible to constrain the range of possible solutions for the CDFs~\cite{Rohrmoser:2018shp}.
We denote $y_0$ as the position of the maximum of $f(y)$, and consider the two extreme cases:
\begin{enumerate}
\item $f(y_0)={\omega}/{2}$: In that case the string-end-point distributions $g_1(y)$ and $g_2(y)$ for both ends of the strings are identical. We label this case as ``$g_1=g_2$''.
Of course, in this case also $G_1(y)=G_2(y)$.
\item $f(y_0)=\omega$: In this case, the supports for the string-end-point distributions $g_1(y)$ and $g_2(y)$ in rapidity are disjoint, hence we refer to this case as the ``disjoint case".
The distribution of the left end-point, $g_1$, has support for $y \le y_0$, whereas the distribution of the right end-point, $g_1$, has support for $y \ge y_0$.
Correspondingly, $G_1(y)=1$ for  $y \ge y_0$ and $G_2(y)=0$ for  $y \le y_0$.
\end{enumerate}

\begin{figure}
\centering
\includegraphics[width=0.46\textwidth]{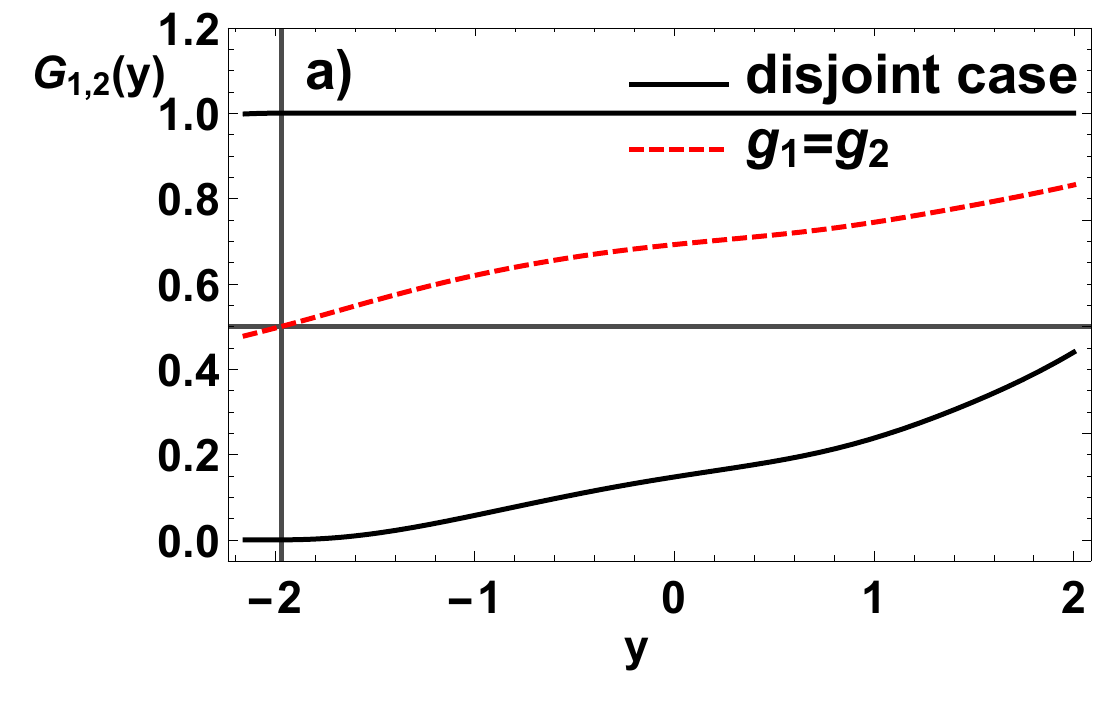}\\
\includegraphics[width=0.46\textwidth]{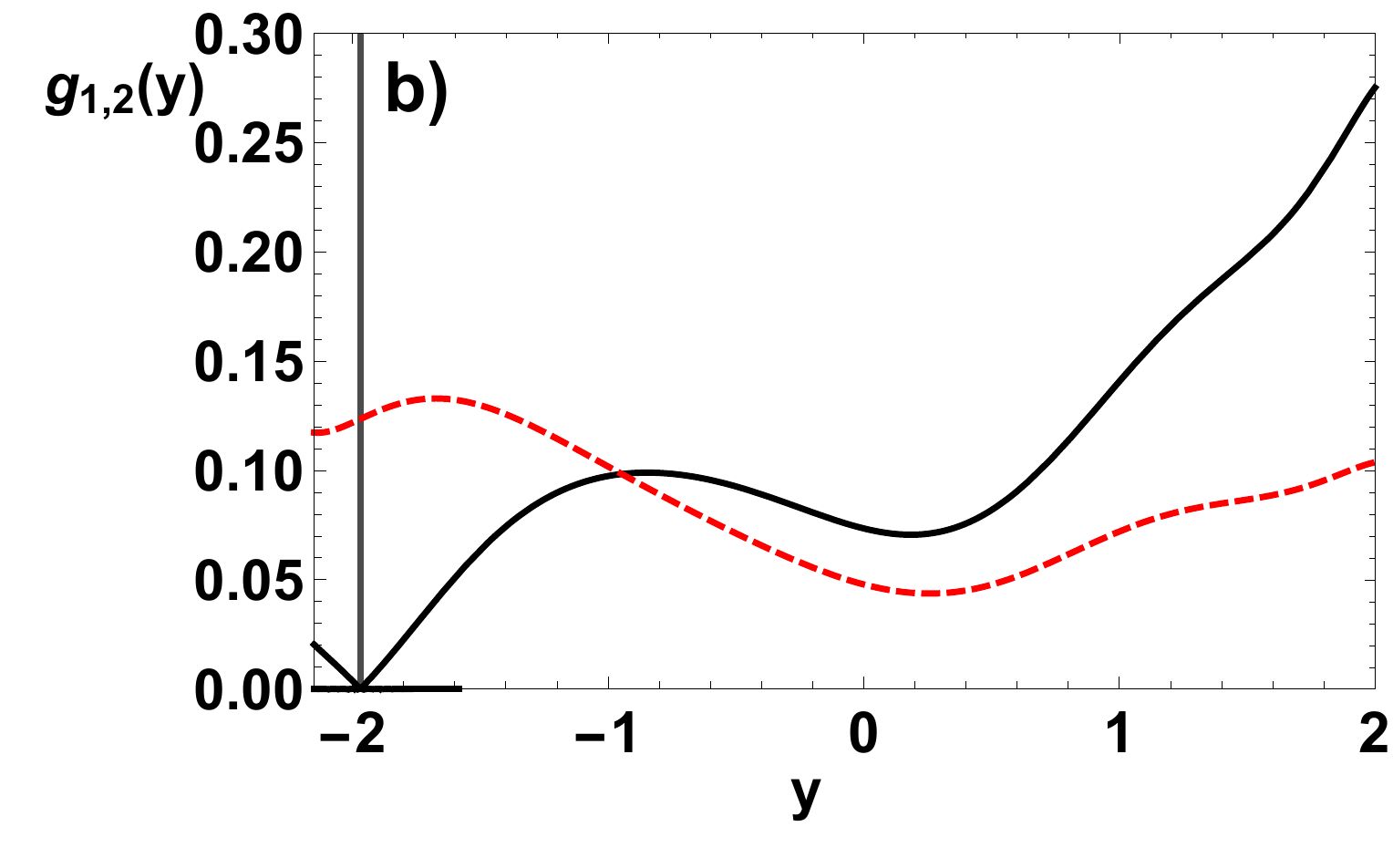}
\caption{Solutions for the cumulative distribution functions of string end-points (a) and the corresponding 
probability distributions (b) in the  $g_1=g_2$ and disjoint cases (dashed and solid lines, respectively). Model with 5 constituent partons per nucleon.
\label{fig:g1g2}}
\end{figure}

Figure~\ref{fig:g1g2}a) shows these two limiting CDFs obtained with the profile $f(y)$ from Fig.~\ref{fig:profiley} for the 5-parton case. 
and Fig.~\ref{fig:g1g2}b) gives the corresponding string end-point distributions, $g_1(y)$ and $g_2(y)$. The position of the maximum is  $y_0 \simeq -2$.
We note the desired features mentioned above. The disjoint case is interpreted in such a way that the left end-point is always at $y \le y_0$, essentially outside of the scope of the plot, whereas 
the right end-point is smoothly distributed at $y \ge y_0$, with highest probability at high values of $y$. 
As discussed in~\cite{Rohrmoser:2018shp}, any solution of Eq.~(\ref{eq:fg1g2}) must have $G_1(y)$ between the 
upper solid line and the dashed line, and  $G_2(y)$ between the 
lower solid line and the dashed line in Fig.~\ref{fig:g1g2}a). This provides useful constraints that carry over to the analysis of two-body correlations.  

\section{Two-particle correlations}

This section presents our model results for the two-particle correlations in pseudo-rapidity obtained for
p-Pb and Pb-Pb collisions at $\sqrt{s_{\rm NN}}=5.02$~TeV.
The findings presented here complement our earlier results~\cite{Rohrmoser:2018shp} 
for d-Au and Au-Au collisions at $\sqrt{s_{\rm NN}}=200$~GeV, with the main difference 
that at $\sqrt{s_{\rm NN}}=5.02$~TeV a wounded parton model with 4 or 5 constituents per nucleon 
is used, rather than the model with 3 constituents per nucleon applied at $\sqrt{s_{\rm NN}}=200$~GeV.

The interesting feature that at higher collision energies one needs in the wounded picture more partons per nucleon has also been
discussed in~\cite{Loizides:2016djv,Bozek:2016kpf} within analyses of the particle multiplicities $dN_{\rm ch}/d\eta$ in A-A collisions. 
Our results are in line with the conclusion of \cite{Loizides:2016djv}, stating that whereas the fits at RHIC collision energies lead to 3 constituents partons, 
higher collision energies prefer about 5 partons per nucleon. 

Two-particle correlations in $A$-$B$ collisions are defined as 
\begin{eqnarray}
C_{AB}(\eta_1,\eta_2) =\frac{\langle N(\eta_1,\eta_2) \rangle}{\langle N(\eta_1) \rangle \langle N(\eta_2) \rangle}, \label{eq:C1}
\end{eqnarray}
where $N(\eta_1,\eta_2)$ is the number of pairs with one particle in a bin centered at $\eta_1$ and the other in a bin centered at $\eta_2$, and $N(\eta_i)$ is the number of 
particles in a bin centered at $\eta_i$. To the extent that $\eta \approx y$ (see the discussion in Sec.~\ref{sec:str}) and 
applying Eqs.~(\ref{eq:yeta},\ref{eq:yeta2}), we may write 
\begin{eqnarray}
C_{AB}(\eta_1,\eta_2) \approx C_{AB}(y_1,y_2), \label{eq:C2}
\end{eqnarray}
since the Jacobian factors $d\eta/dy$ cancel out between the numerator and denominator. 

In analogy to the profile for the emission of individual particles from a single string, a two-particle
profile for the emission of particle pairs from single strings is~\cite{Rohrmoser:2018shp}
\begin{eqnarray}
f_2(y_1,y_2) &=& \omega^2  \label{eq:fnG} G_1[{\rm min}(y_1,y_2)] \left \{1-G_2[{\rm max}(y_1,y_2)] \right \} \nonumber\\
&+& (1 \leftrightarrow 2).
\end{eqnarray}
With this profile one obtains the correlation in pseudorapidity for particle pairs emitted from all strings in $A$-$B$ collisions as 
\begin{eqnarray}
C_{AB}(y_1,y_2) =1+ \frac{{\rm cov}_{AB}(y_1,y_2)}{f(y_1)f(y_2)}, \label{eq:C}
\end{eqnarray}
where ${\rm cov}_{\rm AB}(y_1, y_2)$ is 
\begin{eqnarray}
& {\rm cov}_{\rm AB}&(y_1 , y_2)=  \label{eq:gen} \\
 &\,&\br{N_A} {\rm cov}(y_1,y_2) + \br{N_B} {\rm cov}(-y_1,-y_2)   \nonumber\\
                            &+& {\rm var}(N_A) {f(y_1)}{f(y_2)} + {\rm var}(N_B) {f(-y_1)}{f(-y_2)} \nonumber \\
                            &+& {\rm cov}(N_A,N_B) \left [{f(y_1)} {f(-y_2)}+ {f(-y_1)} {f(y_2)} \right ]. \nonumber
\end{eqnarray}
Contributions to this expression come from emission of a hadron pair from the same string (associated to a wounded parton in $A$ or $B$ nucleus) and from the case where 
the two hadrons originate from different strings. 

\begin{figure}
\centering
\includegraphics[width=0.46\textwidth]{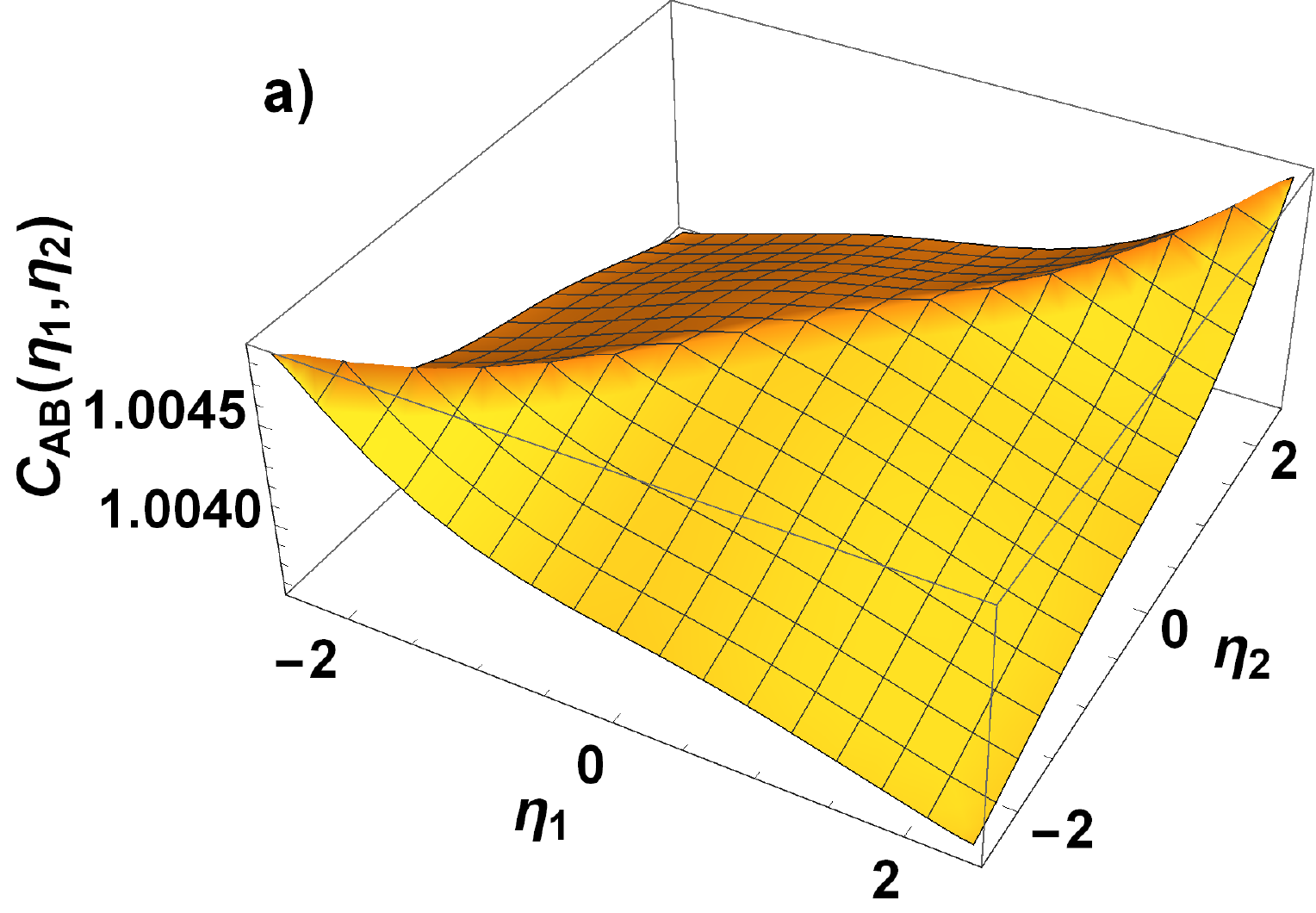}
\includegraphics[width=0.46\textwidth]{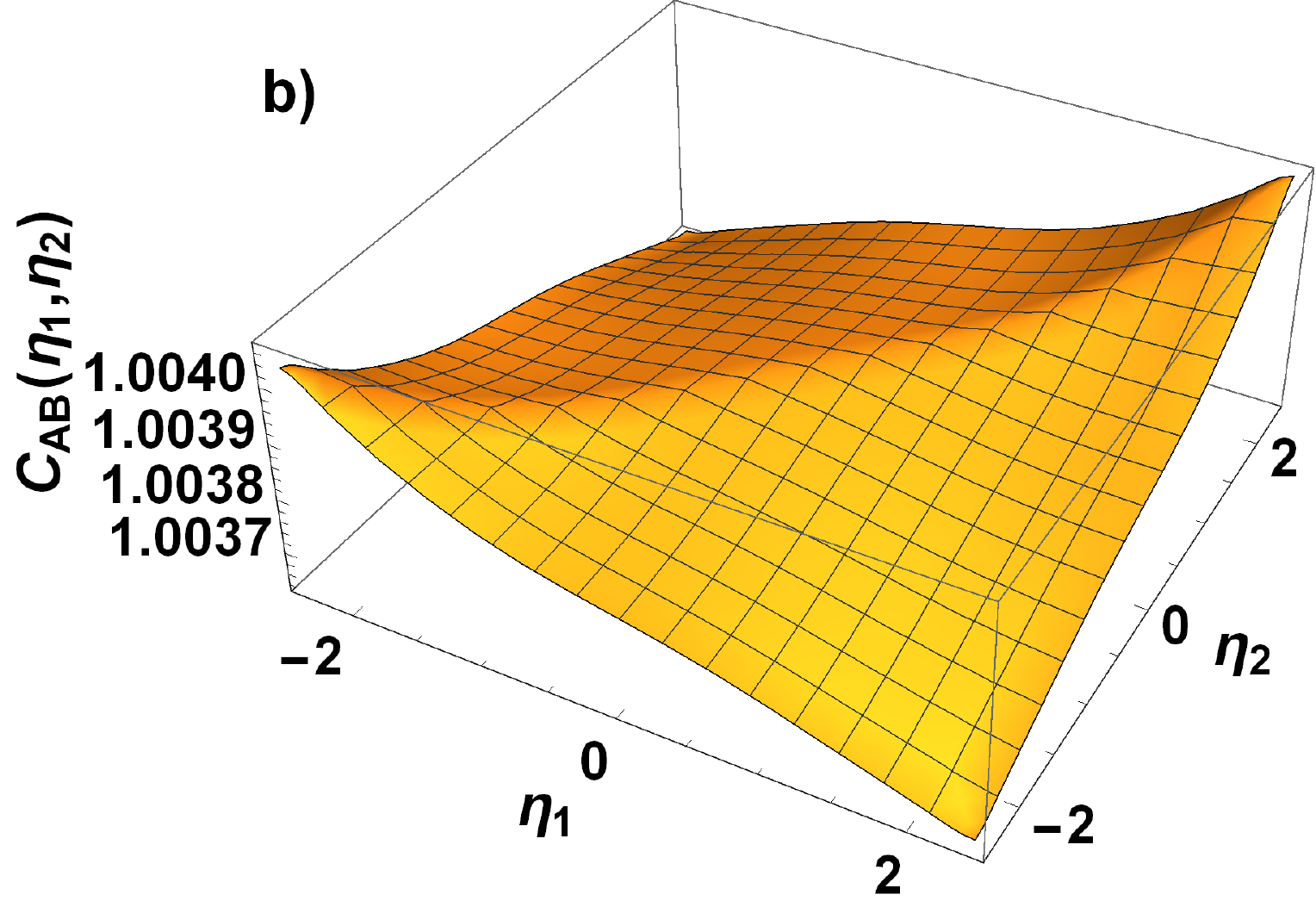}
\includegraphics[width=0.46\textwidth]{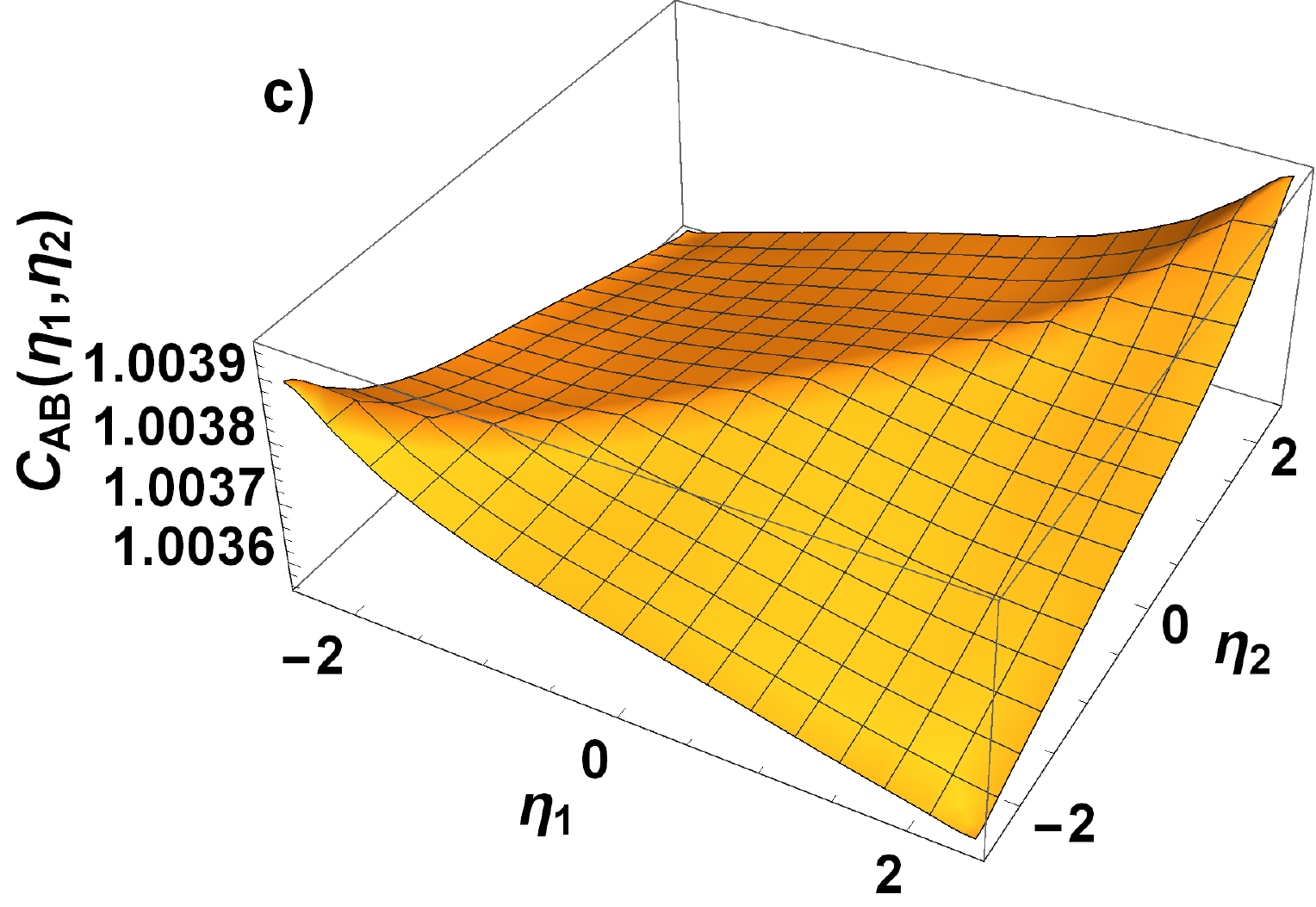}
\caption{Correlations $C(\eta_1,\eta_2)$ in pseudorapidity for the $g_1=g_2$ (a) and the disjoint  (b) cases 
for the $5\%$ most central Pb-Pb collisions at $\sqrt{s_{\rm NN}}=5.02$~TeV for the model with 5 constituent partons per nucleon. 
In (c) the disjoint case for the model with 4 constituent partons per nucleon is shown.}
\label{fig:corr}
\end{figure}

We emphasize that while the different string end-point distributions found in the previous section yield the same one-body 
emission spectra by construction, the same is not in general true for the corresponding two-particle correlations. 
Indeed, noticeable differences occur, as can be seen in Fig.~\ref{fig:corr}, where results 
for $C_{\rm AB}(\eta_1,\eta_2) \simeq C_{\rm AB}(y_1,y_2) $ in both the $g_1=g_2$ and the disjoint cases are shown: 
Both cases yield correlations with a ridge-like structure along the $\eta_1=\eta_2$ direction. 
However for the $g_1=g_2$ case the ridge is higher than that of the disjoint case and, thus, exhibits a steeper decrease in the $\eta_1=-\eta_2$
direction.
We found the same qualitative behavior also for correlations from d-Au and Au-Au collisions at $200$~GeV~\cite{Rohrmoser:2018shp}. 
For comparison, we also show in Fig.~\ref{fig:corr}c) the results for the $4$ constituent model in the disjoint case, 
which is close to the $5$ constituent case from panel~b).

\begin{figure}
\centering
\includegraphics[width=0.46\textwidth]{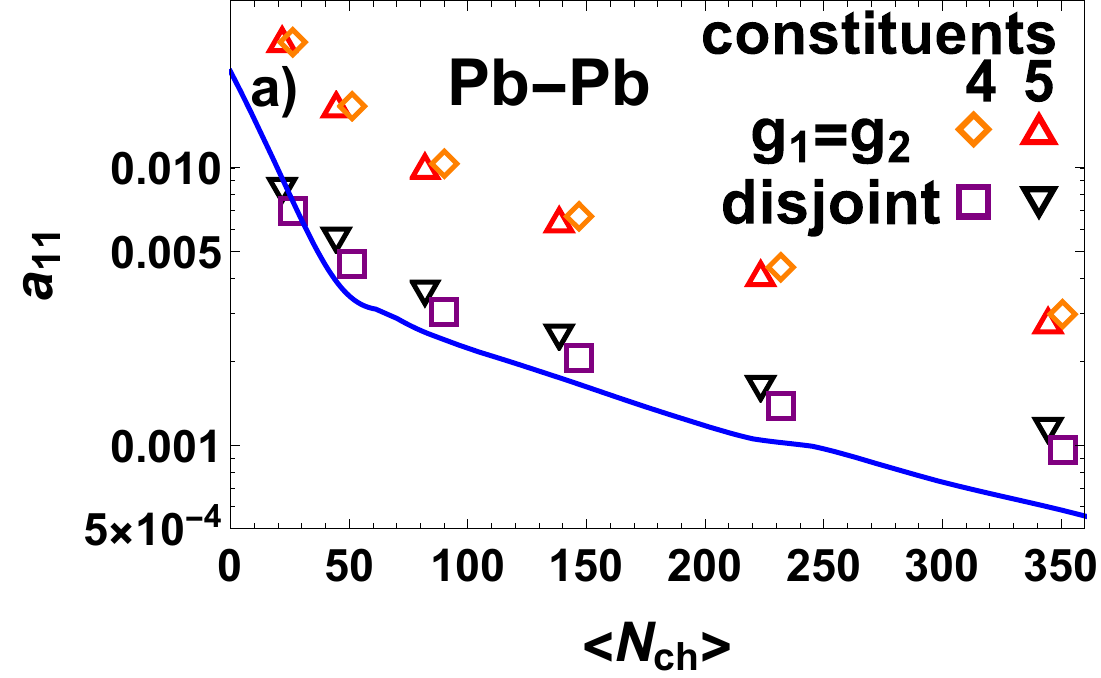}\\
\includegraphics[width=0.46\textwidth]{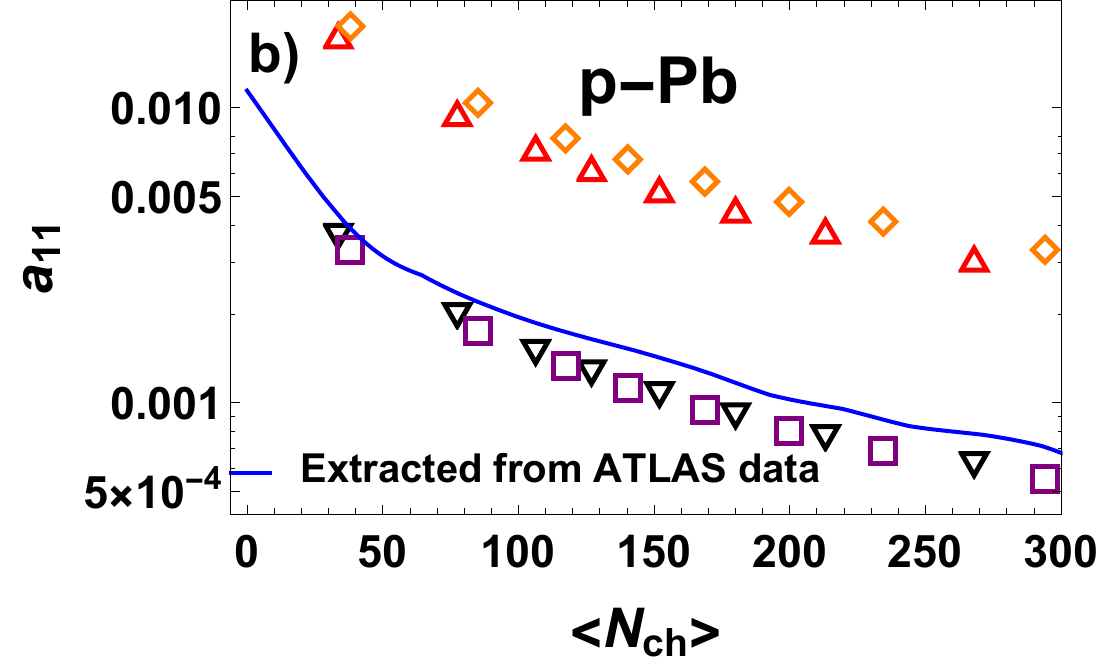}
\caption{Model results for the coefficients $a_{\rm 11}$ for Pb-Pb (a) and p-Pb (b) collisions 
at $\sqrt{s_{\rm NN}}=5.02$~TeV, corresponding to the $g_1=g_2$ and disjoint cases, plotted as functions of the number of charged particles $N_{\rm ch}$ (points), 
in comparison to experimental data from ATLAS~\cite{Aaboud:2016jnr} (solid line) at $\sqrt{s_{NN}}=2.76$~TeV 
and at $\sqrt{s_{NN}}=5.02$~TeV for Pb-Pb and p-Pb collisions respectively. Models with 4 and 5 constituent partons per nucleon.
}
\label{fig:a11}
\end{figure}

To analyze $C_{\rm AB}(y_1,y_2)$ in more quantitative detail, we also study its projections on the Legendre polynomials~\cite{Bzdak:2012tp}
\begin{eqnarray}
a_{nm} &=& \frac{\int_{-Y}^Y {d y_1} \int_{-Y}^Y {d y_2} C(y_1,y_2)
T_n\left(\frac{y_1}{Y}\right) T_m\left(\frac{y_2}{Y}\right)}
{\int_{-Y}^Y {d y_1} \int_{-Y}^Y {d y_2} C(y_1,y_2)}\,, \label{eq:anmC}
\end{eqnarray}
where we follow the choice of $Y=2.4$ of the ATLAS collaboration in order to be able to compare with their results.
The dominant contributions to $C_{\rm AB}(y_1,y_2)$ are represented by the $a_{11}$ coefficients. Our model results for p-Pb 
and Pb-Pb collisions at $\sqrt{s_{\rm NN}}=5.02$~TeV are shown in Fig.~\ref{fig:a11} as a function of the number of charged particles $N_{\rm ch}$ that are produced within the collisions.
As expected, the larger fall-off from the ridge for the correlations in the $g_1=g_2$ case is reflected in larger $a_{11}$ coefficients. 
Our results are shown in comparison to values extracted from ATLAS data for Pb-Pb collisions at $\sqrt{s_{NN}}=2.76$~TeV and p-Pb collisions at $\sqrt{s_{NN}}=5.02$~TeV
from~\cite{Aaboud:2016jnr}. We use the data for $a_{11}$ subtracted by the contribution coming from the short range interactions.

To show our model results as functions of $\langle N_{\rm ch}\rangle$ rather than $\langle N_+\rangle$, we infer from Eq.~(\ref{eq:sym}) that
\begin{equation}
\langle N_{\rm ch}\rangle=\langle N_+\rangle\int_{-Y}^Yd\eta f_s(\eta)\,.
\end{equation}
From this relation we obtain the proportionality  $\langle N_{\rm ch}\rangle=5.76 \langle N_+\rangle$ and $\langle N_{\rm ch}\rangle=5.43 \langle N_+\rangle$ in 
the case of p-Pb and Pb-Pb collisions, respectively (both with $5$ constituents per nucleon). 
For the $4$ constituent model the corresponding values are $\langle N_{\rm ch}\rangle=6.97 \langle N_+\rangle$ and $\langle N_{\rm ch}\rangle=6.64 \langle N_+\rangle$ for p-Pb and Pb-Pb collisions, respectively.

We note from Fig.~\ref{fig:a11} that the model results for  the $a_{11}$ coefficient for the disjoint case are close to the
ATLAS data, compared  to the $g_1=g_2$ case which largely overestimates the data by about a factor of 4. 
We alert the reader that for the Pb-Pb there is a mismatch in the collision energy, as the model analysis is carried for $\sqrt{s_{NN}}=5.02$~TeV, while the data 
are available for $\sqrt{s_{NN}}=2.76$~TeV. Numerically, the mismatch is not significant. 
For comparison, we show in Fig.~\ref{fig:a11} the results for the $4$ and $5$ constituent model, which are very close to each other.

\begin{figure}
\centering
\includegraphics[width=0.46\textwidth]{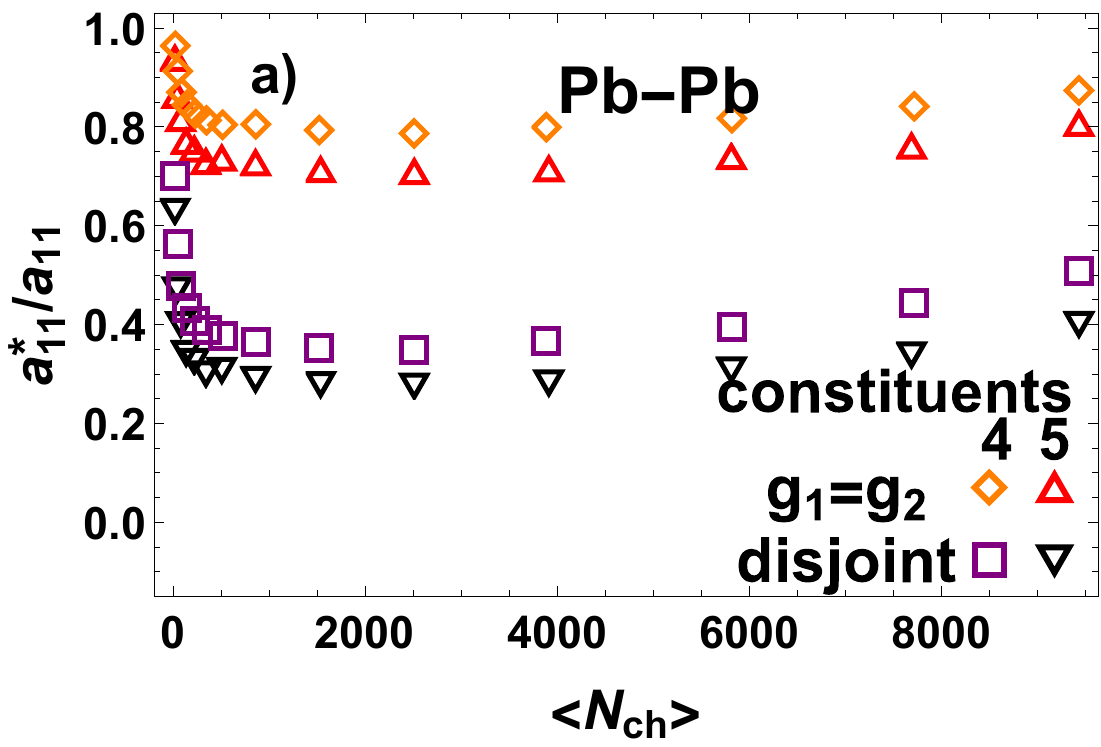}
\includegraphics[width=0.46\textwidth]{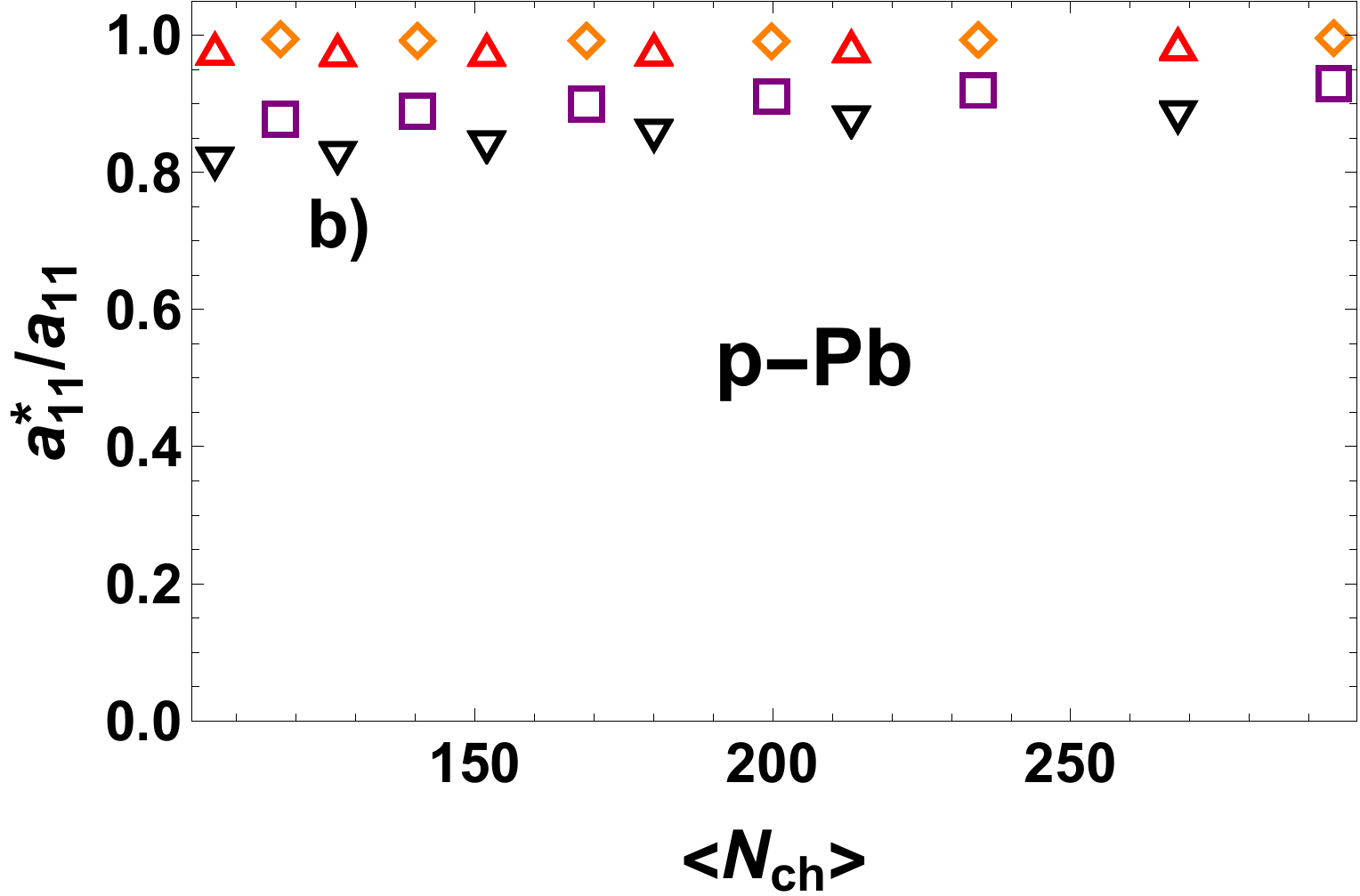}
\caption{Ratios $a^\ast_{11}/a_{11}$ for Pb-Pb (a) and p-Pb (b) collisions 
at $\sqrt{s_{\rm NN}}=5.02$~TeV as functions of the number of charged particles $N_{\rm ch}$. Models with 4 and 5 constituent partons per nucleon.
}
\label{fig:astar}
\end{figure}

We note that the  model results for $a_{11}$ scale approximately as $1/N_{\rm ch} \sim 1/N_+$, as follows from Eqs.~(\ref{eq:C},\ref{eq:gen}). Speaking of the 
decomposition (\ref{eq:gen}), it is interesting to separate the $\br{N_A} {\rm cov}(y_1,y_2) + \br{N_B} {\rm cov}(-y_1,-y_2)$ term originating 
from intrinsic correlations of emission from a string, from the remainder coming from the fluctuation of the number of strings. Following~\cite{Rohrmoser:2018shp},
we denote the corresponding Legendre coefficient as  $a_{11}^\ast$.
Then the ratio  $a_{11}^\ast/a_{11}$ is a measure of the intrinsic correlations compared to the total. 
This ratio is plotted in Fig.~\ref{fig:astar} for the models with 4 and 5 constituents as a function of the number of produced 
charged particles $N_{\rm ch}$, both p-Pb and Pb-Pb collisions.
As can be seen, for the disjoint case which is close to the data, for the Pb-Pb the ratio is around 0.4, indicating a comparable share of 
the contributions from intrinsic string end-point fluctuations and the fluctuation of the number of strings. For the p-Pb case, the corresponding ratio is 
above 0.8, thus the intrinsic fluctuations dominate here.

\section{p-p collisions}

\begin{figure}[t]
\centering
\includegraphics[width=0.46\textwidth]{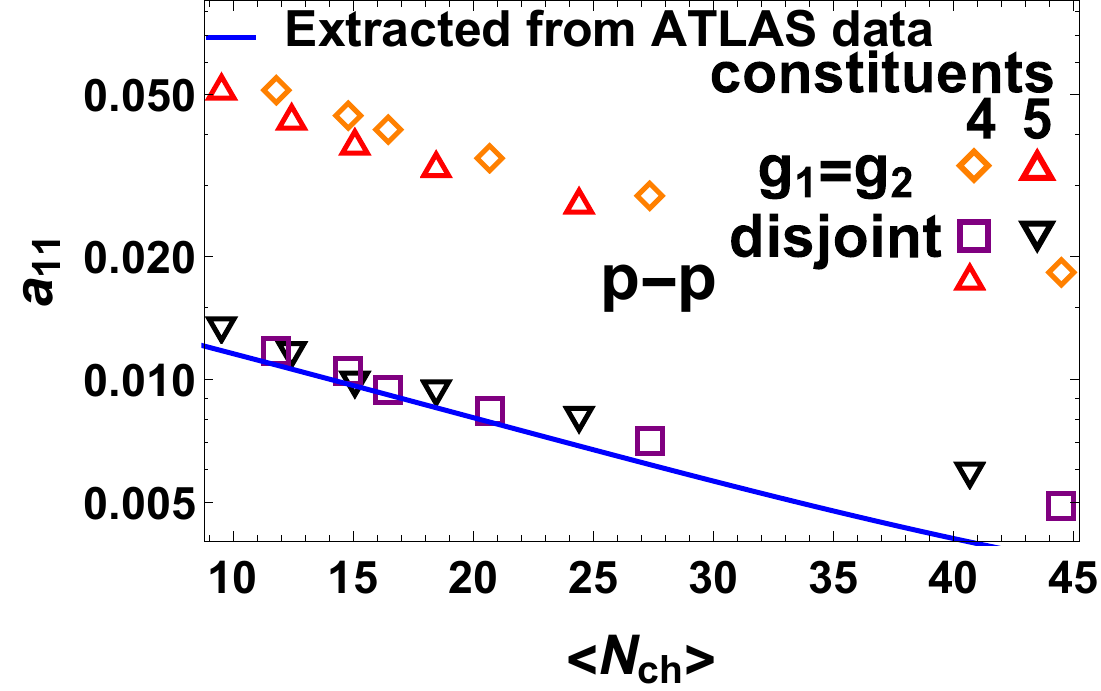}
\caption{The Legendre coefficient $a_{\rm 11}$ for p-p collisions at $\sqrt{s_{\rm NN}}=5.02$~TeV, in 
comparison to the ATLAS data~\cite{Aaboud:2016jnr}  at $\sqrt{s_{\rm NN}}=13$~TeV. Model with 4 and 5 constituent partons per nucleon. \label{fig:pp}}
\end{figure}

With the nucleon substructure present in the model with several constituent partons, it is possible to carry out the correlation analysis also for the p-p collisions. 
In doing so, we use the same emission profile $f(\eta)$, obtained earlier from fitting the Pb-Pb and p-Pb pseudorapidity spectra at $\sqrt{s_{NN}}=5.02$~TeV. 
As before, the numbers of wounded partons are obtained with 
{\tt GLISSANDO~3}~\cite{Bozek:2019wyr} and the negative binomial distribution is overlaid according to Eq.~(\ref{eq:nb}). 
The results, compared to ATLAS data~\cite{Aaboud:2016jnr} at $\sqrt{s_{\rm NN}}=13$~TeV, are presented in Fig.~\ref{fig:pp}.
We note a fair agreement between the model and the experiment, again for the disjoint case. Again, the cases with 4 and 5 partons per nucleon 
are close to each other.

\section{Summary and Conclusions}

The basic conclusion of our study is that a very simple semi-analytic approach involving strings of fluctuating end-points
is capable of explaining the basic features of the long-range two-particle correlation data in pseudorapidity, 
as measured by the ATLAS Collaboration~\cite{Aaboud:2016jnr}. 
In particular, the model with 4 or 5 constituent partons per nucleon and the disjoint distributions for the two fluctuating end-points 
reasonably describes the data for Pb-Pb, p-Pb, and p-p collisions. This explains why more sophisticated models incorporating the 
string breaking mechanism, such as used in various popular Monte Carlo generators, work in describing the longitudinal correlations. 

Our approach merges the wounded constituent model with a generic description of string breaking that was first presented 
in~\cite{Broniowski:2015oif} and~\cite{Rohrmoser:2018shp}, and used for nuclear 
collisions at $\sqrt{s_{NN}}=200$~GeV at RHIC. The extension to the LHC energies, presented here for $\sqrt{s_{NN}}=5.02$~TeV, 
seems phenomenologically successful. Further tests of the model could be performed when the experimental correlation analysis at
other collision energies, broader pseudorapidity coverage, and for other systems become available.

\begin{acknowledgments}
Research supported by the Polish National Science Centre (NCN) Grant 2015/19/B/ST2/00937.
\end{acknowledgments}

\bibliography{hydr}

\end{document}